\documentclass{jpsj2}
\usepackage{multirow}

\usepackage{subfigure}
\newcommand{\url}[1]{\parbox[]{\textwidth}{\tt#1}}

\DeclareRobustCommand\openone{\leavevmode\hbox{\small1\normalsize\kern-.33em1}}

\title{Corpuscular model of two-beam interference and double-slit experiments with single photons}

\author{
\textsc{Fengping Jin}$^{1}$\thanks{E-mail: f.jin@rug.nl},
\textsc{Shengjun Yuan }$^{2}$\thanks{E-mail: s.yuan@science.ru.nl},
\textsc{Hans De Raedt}$^{1}$\thanks{E-mail: h.a.de.raedt@rug.nl},
\textsc{Kristel Michielsen}$^{3}$\thanks{E-mail: k.michielsen@fz-juelich.de}
and
\textsc{Seiji Miyashita}$^{4}$\thanks{E-mail: miya@spin.phys.s.u-tokyo.ac.jp}
}%

\inst{
$^{1}$Department of Applied Physics, Zernike Institute for Advanced Materials,
University of Groningen, Nijenborgh 4, NL-9747AG Groningen, The Netherlands\\
$^{2}$Institute of Molecules and Materials, Radboud University of Nijmegen,
NL-6525ED Nijmegen, The Netherlands \\
$^{3}$Institute for Advanced Simulation, J\"ulich Supercomputing Centre,
Forschungszentrum J\"ulich, D-52425 J\"ulich, Germany \\
$^{4}$
Department of Physics, Graduate School of Science,
The University of Tokyo, 7-3-1 Hongo, Bunkyo-Ku, Tokyo 113-8656
and CREST, JST, 4-1-8 Honcho Kawaguchi, Saitama 332-0012, Japan
}%

\abst{%
We introduce an event-based corpuscular simulation model that reproduces the wave
mechanical results of single-photon double slit and two-beam interference experiments
and (of a one-to-one copy of an experimental realization) of a single-photon interference
experiment with a Fresnel biprism. The simulation comprises models that capture the
essential features of the apparatuses used in the experiment, including the single-photon
detectors recording individual detector clicks. We demonstrate that incorporating in the
detector model, simple and minimalistic processes mimicking the memory and threshold behavior
of single-photon detectors is sufficient to produce multipath interference patterns.
These multipath interference patterns are built up by individual particles taking one
single path to the detector where they arrive one-by-one.
The particles in our model are not corpuscular in the standard, classical physics sense in that they
are information carriers that exchange information with the apparatuses of the experimental set-up.
The interference pattern is the final, collective outcome of the information exchanges
of many particles with these apparatuses.
The interference patterns
are produced without making reference to the solution of a wave equation and without introducing
signalling or
non-local interactions between the particles or between different detection points on the detector screen.
}%

\kword{Computer simulation, Interference, double-slit experiments, quantum theory}

\begin{document}
\maketitle

\section{Introduction}\label{introduction}
In 1802, Young performed a double-slit experiment with light in order to resolve the question whether light was composed of particles,
confirming Newton's particle picture of light, or rather consisted of waves~\cite{YOUNG}.
His experiment showed that the light emerging from the slits produces a fringe pattern on the screen that is characteristic for interference, discrediting
Newton's corpuscular theory of light~\cite{YOUNG}.
Hence, from the point of view of classical physics, the particle and wave character of light did not seem to be compatible.
Moreover, the interpretation in terms of particles or waves of the observations in experiments with light became even more
complicated after conduction of the Michelson-Morley experiment~\cite{MICH87} which provided evidence that light waves
do not need a medium (the ether) to propagate through, in contrast to water and sound waves which require media.
However, explanation of the photoelectric effect by Einstein in terms of photons~\cite{EINS05a} is perhaps the most direct
and convincing evidence of the corpuscular nature of light.
Einstein's explanation of the photoelectric effect was the start of understanding the quantum nature of light
and influenced the development of the concept of wave-particle duality in quantum theory.

In 1924, de Broglie introduced the idea that also matter, not just light, can exhibit wave-like properties~\cite{BROG25}.
This idea has been confirmed in various double-slit experiments with massive objects such as
electrons~\cite{JONS61,MERL76,TONO89,NOEL95}, neutrons~\cite{ZEIL88,RAUC00}, atoms~\cite{CARN91,KEIT91}
and molecules such as $C_{60}$ and $C_{70}$~\cite{ARND99,BREZ02}, all showing interference.
In some of the double-slit experiments~\cite{MERL76,TONO89,JACQ05}
the interference pattern is built up by recording individual clicks of the detectors.
Identifying the registration of a detector click, the ``event'',  with the arrival of a particle
and assuming that the time between successive clicks is sufficiently long such that
these particles do not interact, it becomes a challenge
to explain how the detection of individual objects that do not interact with each other
can give rise to the interference patterns that are being observed.
According to Feynman, the observation that the interference patterns are built up event-by-event
is ``impossible, absolutely impossible to explain in any classical way
and has in it the heart of quantum mechanics''~\cite{FEYN65}.

Although wave-particle duality is a central concept of quantum theory,
in practice quantum theory only works with wave functions to describe the total system under study.
In order to describe the single occurrences observed in various experiments the process of
wave function collapse has been introduced. However, the precise mechanism of a wave function collapse is not known.

Recently, various experiments have been performed that measure individual events generated by microscopic objects.
Hence, it is of interest to study how the particle and wave picture of these experiments are contradicting each other.
It is often said that wave properties like interference cannot be realized by non-interacting particles which satisfy
Einstein's criterion of local causality. In earlier work we have presented an event-based corpuscular simulation model
which demonstrates that such particles can indeed produce interference patterns and applied it to a variety of
single-photon experiments like beam splitter and Mach-Zehnder interferometer experiments,
Wheeler's delayed choice experiments and many
others~\cite{RAED05d,RAED05b,RAED05c,MICH05,RAED06c,RAED07a,RAED07b,RAED07c,ZHAO07b,ZHAO08,ZHAO08b,JIN09c}.
What these experiments have in common is that the interference
can be described as two-path interference, that is the observed interference pattern is the result of having only
two possible paths for the particles travelling between source and detector.
In order to simulate such experiments it is sufficient to use adaptive models for the optical apparatuses
and to use detectors that simply count the number of detection
events~\cite{RAED05d,RAED05b,RAED05c,MICH05,RAED06c,RAED07a,RAED07b,RAED07c,ZHAO07b,ZHAO08,ZHAO08b,JIN09c}.
In this paper we extend the simulation model towards simulating multipath interference patterns
as observed in single-photon two-beam interference and two-slit experiments, for example.
Detectors that are simply counting the detection events cannot be used for this purpose.
Therefore we introduce a new simulation model for the single-photon detector that takes into account
the memory and threshold behavior of such a detector.
The model is a natural extension of the earlier work mentioned and is fully compatible,
that is interchanging in our earlier work the simple counting detector model with this more complex detector
does not change the conclusions. In this sense, the present detector model adds a new, fully compatible,
component to the collection of event-by-event simulation algorithms.

Note that the event-based simulation model is not a corpuscular model in the classical-physics sense.
In our model, particles are objects that carry information. As a particle encounters a material
device, it exchanges information with this device.
In our model, this information exchange is the cause
of the appearance of an interference pattern.
In other words, in our approach we construct a mechanism which produces
wave-like phenomena by local variables only. To this end, we introduce
independent objects which carry information. These objects we call ``particles''.
Each particle interacts with the material of the device only
and the effect of many of such interactions is to build up a
situation which causes the appearance of a first-order interference pattern.

To head off possible misunderstandings, the present paper is not concerned with an
interpretation or an extension of quantum theory
nor does it affect the validity and applicability of quantum theory.
Furthermore, the event-based detector models that we introduce in this paper should not be regarded
as realistic models for say, a photomultiplier or a photographic plate and the
chemical process that renders the image.
Our aim is to show that, in the spirit of Occam's razor, these
very simple event-based models can produce interference patterns
without making reference to the solution of a wave equation.

Although waves can be the physical cause of interference, the key point of our work is that it is wrong to think that
waves are the {\bf only} possible physical cause of interference:
In our approach, the clicks produced by non-interacting / non-communicating detectors, caused
by non-interacting / non-communicating particles that arrive at single detectors one-by-one,
build up a pattern that is identical to the one that is obtained by solving a wave equation.
However, our event-based simulation approach does not require
knowledge of the wave amplitudes obtained by first solving the wave mechanical problem
or requires the solution of the Schr\"odinger equation.
Interference patterns appear as a result of an event-by-event simulation of classical, locally causal, adaptive dynamical systems.

The paper is structured as follows.
In Section~\ref{twobeams}, we introduce the interference experiments that we simulate.
In Section~\ref{event}, we review the main features of the photon detection process.
Section~\ref{model} specifies the new detector models and the simulation algorithm in full detail.
A Mathematica implementation of this algorithm for the case of the double-slit experiment
can be downloaded from the Wolfram Demonstration Project web site~\cite{DS08}.
In Section~\ref{results}, we compare
the event-by-event simulation results with
the numerical results obtained from wave theory for the
two-beam interference experiments discussed in Section~\ref{twobeams},
showing that our event-based, particle-like approach reproduces the
results of quantum theory without making use of concepts thereof.
In Section~\ref{proposal}, we propose a realizable
experiment to test our event-based models for interference.
Our conclusions are given in Section~\ref{conclusion}.

\section{Two-beam interference}\label{twobeams}

\begin{figure}[t]
\begin{center}
\includegraphics[width=10cm]{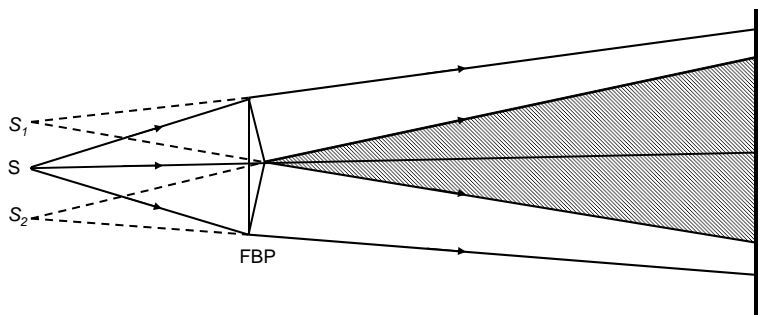}
\caption{Schematic diagram of an interference experiment with a Fresnel biprism (FBP)~\cite{BORN64}.
$S$, $S_1$, $S_2$ denote the point source and its two virtual images, respectively. The grey area is the region
 in which an interference pattern can be observed.}
\label{fig1}
\end{center}
\end{figure}

\begin{figure}[t]
\begin{center}
\includegraphics[width=10cm]{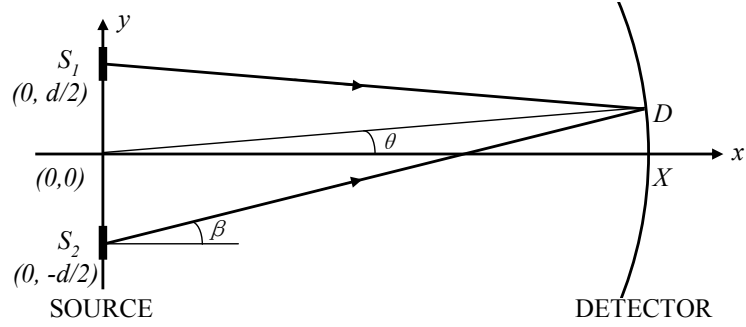}
\caption{Schematic diagram of a simplified double-slit experiment with two sources $S_1$ and $S_2$ of width $a$,
separated by a center-to-center distance $d$,
emitting light according to a uniform current distribution (see Eq.~(\ref{Jy1}))
and with a uniform angular distribution, $\beta$ denoting the angle.
The light is recorded by detectors $D$ positioned on a semi-circle with radius $X$.
The angular position of a detector is denoted by $\theta$.
}
\label{fig2}
\end{center}
\end{figure}

\begin{figure}[t]
\begin{center}
\includegraphics[width=10cm]{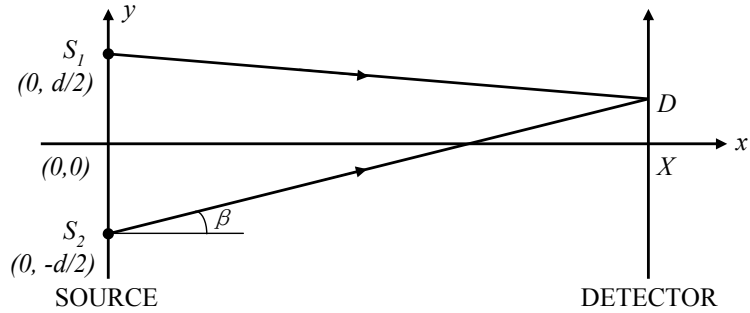}
\caption{Schematic diagram of a two-beam interference experiment with two line sources $S_1$ and $S_2$ having a spatial
Gaussian profile (see Eq.~(\ref{Jy})), emitting light according to a uniform angular distribution, $\beta$ denoting the angle.
The sources are separated by a center-to-center distance $d$.
The light is detected by detectors $D$ positioned at $(X,y)$.
}
\label{fig3}
\end{center}
\end{figure}

In this paper, we focus on interference experiments with single-photons, leaving the case
of massive particles for further research.
As a prototype problem, we consider two-beam interference experiments with a Fresnel biprism~\cite{BORN64}.
A schematic diagram of such an experiment is shown in Figure~\ref{fig1}.
A pencil of light, emitted by the source $S$, is divided by refraction into two pencils~\cite{BORN64}.
Interference can be obtained in the region where both pencils overlap, denoted by the grey area in Fig.~\ref{fig1}.
As a Fresnel biprism consists of two equal prisms with small refraction angle and as the angular aperture of the pencils is small,
we may neglect aberrations~\cite{BORN64}.
The system consisting of the source $S$ and the Fresnel biprism can then be replaced by
a system with two virtual sources $S_1$ and $S_2$~\cite{BORN64}, see Fig.~\ref{fig1}.
Alternatively, following Young~\cite{BORN64} we can let the light impinge on a screen
with two apertures and regard these apertures as the two virtual sources $S_1$ and $S_2$, see Figs.~\ref{fig2} and \ref{fig3}.
Results of a single-photon interference experiment with a Fresnel biprism
and a time-resolved interference experiment for the system schematically depicted in Fig.~\ref{fig3} are reported
in Refs.~\cite{JACQ05} and \cite{SAVE02}, respectively.

For all these simplified systems, a straightforward application of Maxwell's theory yields the intensity at the
detection screen.
We consider a few representative cases for which closed-form expressions can be obtained:
\begin{itemize}
\item{The sources $S_1$ and $S_2$ are lines of length $a$, separated by a center-to-center distance $d$, see Fig.~\ref{fig2}.
These sources emit light according to a uniform current distribution, that is
\begin{equation}
J(x,y)=\delta(x)\left[\Theta(a/2-|y-d/2|)+\Theta(a/2-|y+d/2|)\right]
,
\label{Jy1}
\end{equation}
where $\Theta(.)$ denotes the unit step function.
In the Fraunhofer regime, the light intensity at the detector on a circular screen is given by~\cite{BORN64}
\begin{equation}
I(\theta) = A\left(\frac{\sin\frac{qa\sin\theta}{2}}{\frac{qa\sin\theta}{2}}\right)^2 \cos^2\frac{qd\sin\theta}{2},
\label{eq_slit}
\end{equation}
where $A$ is a constant, $q$ is the wave number,
and $\theta$ denotes the angular position of the detector $D$ on the circular screen, see Fig.~\ref{fig2}.
}
\item{The sources $S_1$ and $S_2$ form a line source with a current distribution given by
\begin{equation}
 J(x,y) = \delta(x)\sum_{s=\pm1} e^{-(y-sd/2)^2/2\sigma^2}
,
\label{Jy}
\end{equation}
where $\sigma$ is the variance and $d$ denotes the distance between the centers of the two sources, see Fig.~\ref{fig3}.
The intensity of the overlapping pencils at the detector reads
\begin{equation}
I(y) = B\left(\cosh\frac{byd}{\sigma ^2} + \cos\frac{(1-b)qyd}{X} \right)e^{-b(y^2+d^2/4)/\sigma ^2},
\label{eq_Gaussian}
\end{equation}
where $B$ is a constant, $b=q^2\sigma^4/(X^2 + q^2\sigma^4)$, and $(X,y)$ are the coordinates
of the detector $D$ (see Fig.~\ref{fig3}).
Closed-form expression Eq.~(\ref{eq_Gaussian}) was obtained by assuming that $d\ll X$ and $\sigma\ll X$.
}
\item{The two sources $S_1$ and $S_2$ are circles with a radius $a$ and their centers are separated by a distance $d$.
The current distribution is given by
\begin{equation}
 J(x,y,z) = \delta(x)\left[\Theta(a^2/4-(y-d/2)^2-z^2)+\Theta(a^2/4-(y+d/2)^2-z^2)\right]
.
\label{Jy2}
\end{equation}
In the Fraunhofer regime, the light intensity at a detector placed on a sphere is given by~\cite{BORN64}
\begin{equation}
I(\theta) = C \left(\frac{2J_1(q a \sin\theta)}{qa \sin\theta}\right)^2 \cos^2\frac{q d \sin\theta}{2}
,
\label{circularslits}
\end{equation}
where $C$ is a constant,
$\theta$ denotes the zenith of the detector $D$ on the spherical detection screen
and $J_1(.)$ is the Bessel function of the first kind of order one.
}
\end{itemize}
From Eqs.~(\ref{eq_slit}), (\ref{eq_Gaussian}) and (\ref{circularslits}),
it directly follows that the intensity distribution on the detection screen,
displays fringes that are characteristic for interference.

\section{Event-by-event simulation and detector model}\label{event}

Imagine that individual particles build up the interference pattern one by one
and exclude the possibility that there is direct communication between the particles
(even if one particle has arrived at the detector while another particle is at the
source or at a detector).
If we then simply look at Fig.~\ref{fig2} or \ref{fig3}, we arrive at the logically unescapable
conclusion that
the interference pattern can only be due to the internal operation of the detector:
There is nothing else that can cause the interference pattern to appear.

Obviously a simple, passive detector model that only counts the number of particles fails to reproduce the interference patterns of
two-beam interference experiments in which there are sources and detectors only, as in Figs.~\ref{fig2} and \ref{fig3}.
Before we introduce new event-based models for the detector,
it is expedient to review the conventional theory of the photon detection process.

In its simplest form, a light detector consists of a material that can be ionized by light.
The electric charges that result from the ionization process are then amplified,
chemically in the case of a photographic plate or electronically in the case
of photo diodes or photomultipliers.
In the wave-mechanical picture,
the interaction between the incident electric field ${\mathbf E}$ and
the material takes the form ${\mathbf P}\cdot{\mathbf E}$, where ${\mathbf P}$ is the polarization vector of the material~\cite{BORN64}.
Treating this interaction in first-order perturbation theory, the detection probability reads
$P_{detection}(t)=\int^{t}_{0}\int^{t}_{0}\langle\langle{\mathbf E^{T}(t')}\cdot{\mathbf K(t'-t'')}\cdot{\mathbf E(t'')} \rangle\rangle dt'dt''$
where ${\mathbf K(t'-t'')}$ is a memory kernel that is characteristic for the material only and
$\langle\langle.\rangle\rangle$ denotes the average with respect to the initial state of the electric field~\cite{BALL03}.
Both the constitutive equation~\cite{BORN64} ${\mathbf P}(\omega)=\chi(\omega){\mathbf E}(\omega)$ as well as the expression
for $P_{detection}(t)$ show that the detection process involves some kind of memory.
Furthermore, very sensitive photon detectors such as photomultipliers and avalanche diodes are
trigger devices, meaning that the recorded signal depends on an intrinsic threshold.
Conceptually, the chemical process that renders the image encoded in the photographic material plays a similar role.

From these general considerations, it is clear that a minimal model for the detector should be able
to account for the memory and the threshold behavior of the detectors.
An event-based model for the detector cannot be ``derived'' from quantum theory,
simply because quantum theory has nothing to say about individual events~\cite{HOME97}.
Therefore, from the perspective of quantum theory,
any model for the detector that operates on the level of single events
must necessarily appear as ``ad hoc''.
In contrast, from the viewpoint of a contextual description, the
introduction of such a model is a necessity~\cite{HOME97}.

\section{Simulation model}\label{model}
In our simulation model, every essential component of the laboratory experiment such as
the source, the Fresnel biprism, and detector array has a counterpart in the algorithm.
The data is analyzed by counting detection events, just as in the laboratory experiment~\cite{JACQ05}.
The simulation model is solely based on experimental facts.

The simulation can be viewed as a message-processing and message-passing process routing messengers
through a network of units that processes messages. The processing units play the role of the components
of the laboratory experiment and the network represents the complete experimental set-up.
We now specify the operation of the basic components of the simulation model in full detail.
Other components that are specific to a particular interference experiment are described
together with the presentation of the simulation results.

\subsection{Messenger}
In our simulation approach, we view each photon as a messenger carrying a message.
Each messenger has its own internal clock, the hand of which rotates with frequency $f$.
As the messenger travels from one position in space to another,
the clock encodes the time-of-flight $t$ modulo the period $1/f$.
The message, the position of the clock's hand, is most conveniently represented by a two-dimensional unit vector
${\mathbf e}_k=(e_{0,k}, e_{1,k})=(\cos\phi_k, \sin\phi_k)$, where
the subscript $k>0$ labels the successive messages,
$\phi_k=2\pi f t_k$, and $t_k$ is the time-of-flight of the $k$-th messenger.
The messenger travels with a speed $c/n$ where $c$ denotes the speed of light in vacuum and
$n$ is the refractive index of the medium in which the messenger moves.

\subsection{Source}
In a simulation model in which the photons are viewed as messengers,
the single-photon source is trivially realized by
creating a messenger and waiting until its message has been processed by the detector before creating
the next messenger.
This ensures that there can be no direct information exchange between the messengers,
even if one particle has arrived at the detector while another particle is at the
source or at a detector,
implying that our simulation model (trivially) satisfies Einstein's criterion of local causality.

For the double-slit, two-beam interference, and circular slits simulations,
messengers leave the source at positions generated randomly
according to the current distributions Eqs.~(\ref{Jy1}), (\ref{Jy}), and (\ref{Jy2}),
respectively. The distribution of the angle $\beta$ is chosen to be uniform.
When messenger $k$ is created, its internal clock time $t_k$ is set to zero.

\subsection{Detector}
\begin{figure}[t]
\begin{center}
\includegraphics[width=10cm]{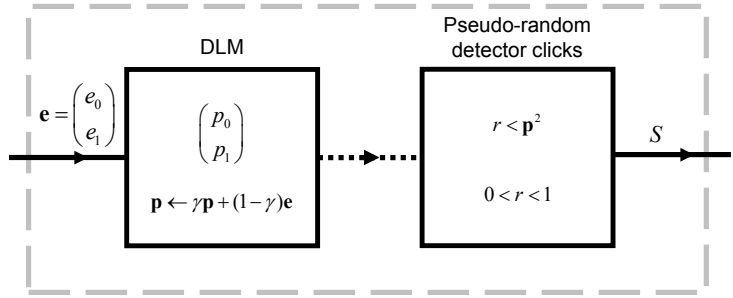}
\caption{%
Diagram of the event-based detector model defined by Eqs.~(\ref{ruleofLM}) and (\ref{thresholdofLM}).
The dashed line indicates the data flow within the processing unit.
}%
\label{fig0}
\end{center}
\end{figure}
A single photon detector, such as a photographic plate, consists of many identical detection units
each having a predefined spatial window in which they can detect photons. Because these small detection
units are photon detectors themselves we also name them detectors in what follows.
Here we construct a processing unit that acts as a detector for individual messages.
A schematic diagram of the unit is shown in Fig.~\ref{fig0}.
The first stage consists of a deterministic learning machine (DLM)
that receives on its input channel the $k$th message represented by
the two-dimensional vector ${\mathbf e}_k=(\cos \phi_k,\sin\phi_k)$.
In its simplest form the DLM contains a single
two-dimensional internal vector with Euclidean norm less or equal than one.
We write ${\mathbf p}_k = (p_{0,k},p_{1,k})$ to denote
the value of this vector after the $k$th message has been received.
Upon receipt of the $k$th message the internal vector is updated according to the rule
\begin{equation}
	{\mathbf p}_k = \gamma {\mathbf p}_{k-1} + (1-\gamma) {\mathbf e}_k,
	\label{ruleofLM}
\end{equation}
where $0<\gamma<1$ and $k>0$.
Update rule Eq.~(\ref{ruleofLM}) clearly indicates that the first stage learns from the incoming messages
in a deterministic way and therefore it is given the name deterministic learning machine.
Obviously, if $\gamma\not=0$, a machine  that operates
according to the update rule Eq.~(\ref{ruleofLM}) has memory.

The second stage of the detector (see Fig.~\ref{fig0}) uses the information
stored in the internal vector to decide whether or not to generate a click (threshold behavior).
As a highly simplified model for the bistable character of the real photodetector or
photographic plate, we let the machine generate a binary output signal $S_k$ using the intrinsic threshold function
\begin{equation}
	S_k = \Theta({\mathbf p}^2_k-r_k),
	\label{thresholdofLM}
\end{equation}
where $\Theta(.)$ is the unit step function and $0\leq r_k <1$ is a uniform pseudo-random number.
Note that in contrast to experiment, in a simulation, we could register both the $S_k=0$ and $S_k=1$ events
such that the number of input messages equals the sum of the $S_k=0$ and $S_k=1$ detection events.
Since in experiment it cannot be known whether a photon has gone undetected, we discard the information about
the $S_k=0$ detection events in our future analysis.

The total detector count is defined as
\begin{equation}
	N=\sum^{k}_{j=1}S_j,
	\label{N_counts}
\end{equation}
where $k$ is the number of messages received. Thus, $N$ counts the number of one's generated by the machine.
As noted before
a detector screen is just a collection of identical detectors and is modeled as such.
Each detector has a predefined spatial window within which it accepts messages.

In Appendix A we prove that as $\gamma\rightarrow1^-$, the internal vector ${\mathbf p}_{k}$
converges to the average of the messages ${\mathbf e}_{1},\ldots,{\mathbf e}_{k}$.
In general, the parameter $\gamma$ controls the precision
with which the machine defined by Eq.~(\ref{ruleofLM}) learns the average of the sequence of messages
${\mathbf e}_{1},{\mathbf e}_{2}, \ldots$
and also controls the pace at which new messages affect the internal state of the DLM (memory effect)~\cite{RAED05d}.
In Appendix B we show how to modify the update rule Eq.~(\ref{ruleofLM})
such that the transient regime of the detector becomes shorter.
The transient behavior of the simplest and the slightly more complicated detector
models may be accessible to real experiments, as explained in Section~\ref{proposal}.
In appendix B, we also give an alternative for Eq.~(\ref{thresholdofLM}) that does not make use
of pseudo-random numbers.

Before we proceed we make a few notes on the memory and threshold behavior of our detector simulation models.
Although the word memory may give the impression that the detector keeps track of all the photons that pass,
all the event-based detector models introduced in this paper have
barely enough memory to store the equivalent of one message.
Thus, these models derive their power, not from storing a lot of data,
but from the way they process successive messages.
Most importantly, the DLMs do not need to keep track of the number $k$ of messages
that they receive, a number that we cannot assume to be known because
in real experiments we can only count the clicks of
the detector, not the photons that were not detected.
As shown in Appendix C, the role of the local memory of the detector is similar to that of the
dielectric function in Maxwell's theory.
Our detector models do not incorporate a memory fade-out as a function of time.
Although this could be an essential feature in time frames in which the detectors do not receive photons,
we do not consider it to be of importance for our present study.

We also want to emphasize that the presence of a threshold does not cause our detector model
to operate with less than 100\% efficiency.
In general, the detection efficiency is defined as the overall probability of registering a count if a
photon arrives at the detector~\cite{HADF09}
Using this definition, our event-based detector model simulates an ideal single-photon detector
that has 100\% detection efficiency.
This can easily be demonstrated by performing the simulation of an experiment
(which is very different from a double-slit experiment)
that measures the detection efficiency~\cite{HADF09}.
In such an experiment a point source emitting single photons (messengers) is placed far away
from a single detector.
As all photons that reach the detector have the same time-of-flight (to very good approximation),
all the messengers that arrive at this detector will carry the same message.
As a result, the internal vector rapidly converges to one,
so that the detector clicks every time a photon arrives.
Thus, the detection efficiency, as defined for real detectors,
of our detector model is very close to 100\%.
Although the detection efficiency of the detector model itself is very close to 100\%,
the ratio of detected to emitted photons is much less than one.
Note however that, in general, as is well known, a photon detector + electronics
is an open system (powered by external electrical sources etc.),
hence photon-energy conservation within the detector-photon system is not an issue.

\subsection{Discussion}
In our approach, interference appears as a result of processing individual events,
but definitely not because we have introduced ``wave-like'' ingredients in a sneaky manner.
In our corpuscular model, each particle carries its own clock, that is, it carries its own local oscillator.
This oscillator only serves to mimic the frequency of the individual particle (photon).
A the particle hits the detector, the detector "observes" the state of the oscillator
that is attached to this particular particle and determines its time-of-flight.
Note that the idea of introducing the time-of-flight does not mean that we obtain
interference by summing wave functions $a_ke^{-i\omega t_k}$ where $t_k$ denotes the time-of-flight of the $k$th particle.

There is no communication/interaction between the detectors that make up the detection screen,
hence there is no wave equation (i.e. no partial differential equation)
that enforces a relation between the internal variables of these detectors.
Likewise, the oscillator that is carried by a particle never interacts with an oscillator
of another particle, hence the motion of these two oscillators is also not governed by a wave equation.
Naively, one might imagine the oscillators tracing out a wavy pattern in space
as they travel from the source to the detector screen.
However, in our model there is no relation between the times at which the particles leave the source,
hence it is impossible to characterize all these traces by a field that
depends on one set of space-time coordinates, as required for a wave theory.

\section{Simulation results}\label{results}

First, we demonstrate that our event-by-event simulation model reproduces the wave mechanical results
Eq.~(\ref{eq_slit}) of the double-slit experiment.
Second, we simulate a two-beam interference experiment
and show that the simulation data agree with Eq.~(\ref{eq_Gaussian}).
Third, we validate the simulation approach by reproducing the interference patterns for two circular
sources, see Eq.~(\ref{circularslits}). Finally, we present the results for the simulation of the single-photon interference experiment with a Fresnel biprism~\cite{JACQ05},
see Fig.~\ref{fig1}.
The results presented in this section have all been obtained using the detector model described in Section~\ref{model}.
Simulation data produced by the detector models described in Appendix B are given in Section~\ref{proposal}.

\begin{figure}[t]
\begin{center}
\includegraphics[width=10cm]{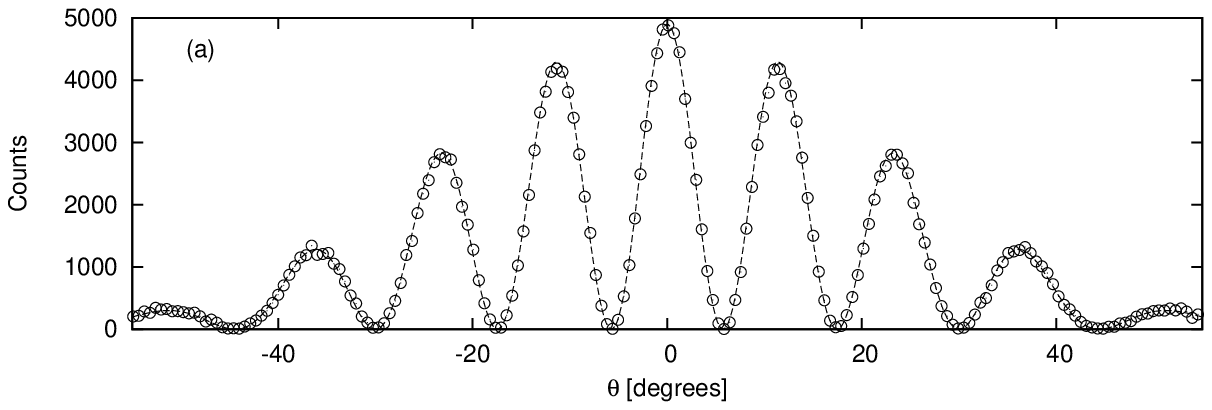}
\includegraphics[width=10cm]{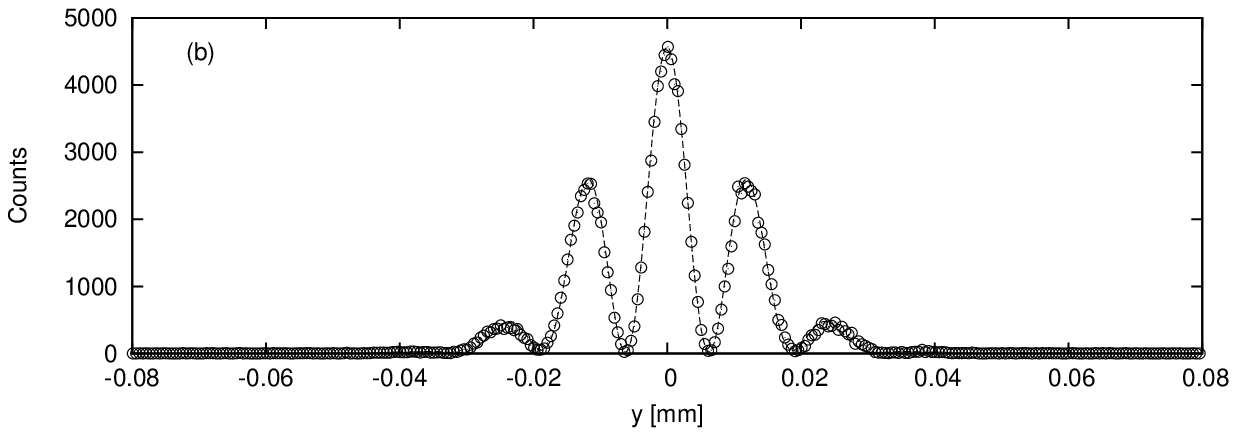}
\includegraphics[width=10cm]{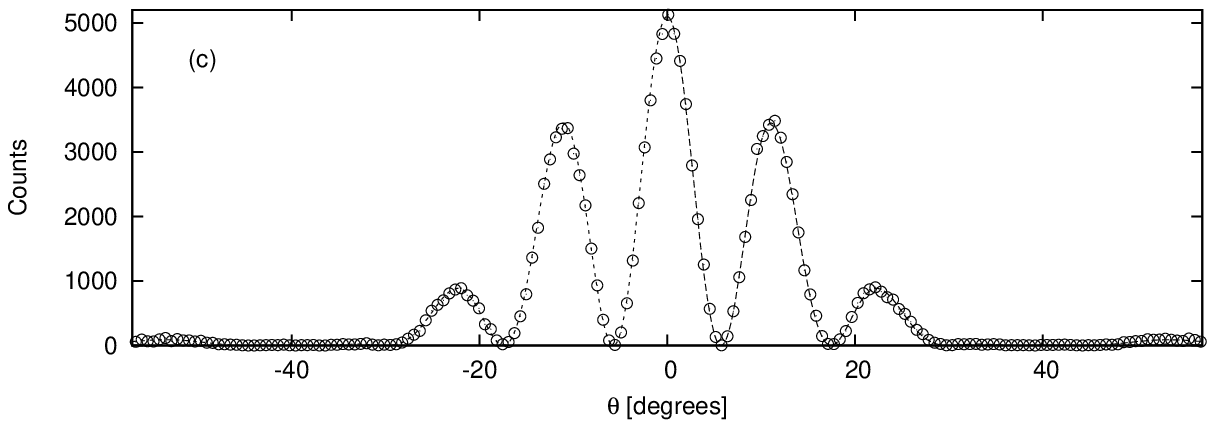}
\caption{Detector counts as a function of the angular (spatial) detector position $\theta$ ($y$)
as obtained from event-by-event simulations of the interference experiment shown in Fig.~\ref{fig2} (Fig.~\ref{fig3}).
The circles denote the event-based simulation results produced by the
detector model defined in Section~\ref{model}.
The dashed lines are the results of wave theory (see Eqs.~(\ref{eq_slit}), (\ref{eq_Gaussian})
and (\ref{circularslits})).
(a) The sources are slits of width $a=\lambda$ ($\lambda=670$ nm in all our simulations),
separated by a distance $d=5\lambda$ and the source-detector distance $X=0.05$ mm, see Fig.~\ref{fig2}.
The sources emit particles according to the current distribution Eq.~(\ref{Jy1}).
An interactive program for the double-slit simulation can be downloaded from the Wolfram Demonstration Project
web site~\cite{DS08};
(b) The sources $S_1$ and $S_2$, separated by a distance $d=8\lambda$, emit particles according
to a Gaussian current distribution Eq.~(\ref{Jy})
with variance $\sigma=\lambda$ and mean $d/2$ and $-d/2$, respectively (see Fig.~\ref{fig3}).
The source-detector distance $X=0.1$ mm;
(c) The two circular sources $S_1$ and $S_2$ of radius $a=\lambda$ with centers separated by a distance $d=5\lambda$
emit particles according to the current distribution Eq.~(\ref{Jy2}).
The distance between the center of the two-source system and the spherical detection screen is $X=0.1$ mm.
}
\label{simu1}
\end{center}
\end{figure}

\subsection{Double-slit experiment}

As a first example, we consider the two-slit experiment with sources that are slits of width $a=\lambda$
($\lambda=670$ nm in all our simulations), separated by a center-to-center distance $d=5\lambda$,
see Fig.~\ref{fig2}.
In Fig.~\ref{simu1}(a), we present the simulation results for a source-detector distance $X=0.05$ mm.
When a messenger (photon) travels from the source at $(0,y)$ to the circular detector screen with radius $X$,
it updates its own time-of-flight, or equivalently its angle $\phi$.
This time-of-flight is calculated according to geometrical optics~\cite{BORN64}.
More specifically, a messenger leaving the source at $(0, y)$ under an angle $\beta$ (see Fig.~\ref{fig2})
will hit the detector screen at a position determined by the angle $\theta$ given by
\begin{equation}
\sin\theta=z\cos^2\beta+\sin\beta\sqrt{1-z^2\cos^2\beta}
,
\end{equation}
where $z=y/X$ and $|z|< 1$.
The distance traveled is then given by
\begin{equation}
s=X\sqrt{1-2z\sin\beta+z^2}
,
\end{equation}
and hence the message is determined by the angle $\phi=2\pi fs/c$ where $c$ is the speed of light.
As the messenger hits a detector, the detector updates its internal vector and decides whether to output a zero or a one.

This process is repeated many times.
The initial $y$-coordinate of the messenger is chosen randomly from
a uniform distribution on the interval $[-d/2-a/2,-d/2+a/2]\cup[+d/2-a/2,+d/2+a/2]$.
The angle $\beta$ is a uniform pseudo-random number between $-\pi/2$ and $\pi/2$.

The markers in Fig.~\ref{simu1}(a) show the event-by-event simulation results
produced by the detector model described in Section~\ref{model} with $\gamma=0.999$.
We used a set of thousand detectors positioned equidistantly in the interval $[-57^{o},57^{o}]$,
each of them receiving on average by six thousand photons.
The number of clicks generated by the detectors, that is the number of so-called detected photons, is approximately $16.10^5$.
Hence, the ratio of detected to emitted photons is of the order 0.25, a fairly large number compared to those achieved
in laboratory experiments with single-photons (see Section~\ref{biprism}).
The result of wave theory, as given by the closed-form expression Eq.~(\ref{eq_slit}), is represented by the dashed line.
Without using any knowledge about the solution of a wave equation,
the event-based simulation (markers) reproduces the results of wave theory.

According to our mathematical analysis of the performance of the machines (see Appendix A),
accurate results (relative to the predictions of quantum theory) are to be expected for $\gamma$ close to one only.
Taking for instance $\gamma=0.99$ does not change the qualitative features
although it changes the number of counts by small amounts (data not shown).

An interactive Mathematica program of the event-based double-slit simulation
which allows the user to change the model parameters and to verify
that the simulation reproduces the results of wave theory
may be downloaded from the Wolfram Demonstration Project web site~\cite{DS08}.

\subsection{Two-beam interference experiment}
As a second example we consider the two-beam interference experiment depicted in Fig.~\ref{fig3}.
We assume that the messengers leave either source $S_1$ or $S_2$ from a position $y$ that is
distributed according to a Gaussian distribution with variance $\sigma$ and mean $+d/2$ or $-d/2$, respectively.
Also in this case, the time-of-flight is calculated according to geometrical optics~\cite{BORN64}.
A messenger leaving the source at $(0, y)$ under an angle $\beta$ (see Fig.~\ref{fig3})
will hit the detector screen at a position $(X, y')$
\begin{equation}
y'=X\tan\beta+y
,
\end{equation}
the distance traveled is given by $s=X\sec\beta$ and
the message is determined by the angle $\phi=2\pi fs/c$ where $c$ is the speed of light.

The simulation results for a source-detector distance $X=0.1$ mm, for $\gamma=0.999$
are shown in Fig.~\ref{simu1}(b).
The dashed line is the result of wave theory, see closed form expression Eq.~(\ref{eq_Gaussian}).
Also in this case, the agreement between wave theory and the event-by-event simulation is extremely good.

\subsection{Double-slit experiment with circular sources}

As a third example, we consider the double-slit experiment with circular sources,
a straightforward extension of the two-dimensional double-slit system to three dimensions.
As shown in Fig.~\ref{simu1}(c), there is excellent agreement between the event-by-event simulation
and the analytical expression Eq.~(\ref{circularslits}).

\subsection{Experiment with a Fresnel biprism}\label{biprism}

\begin{figure}[t]
\begin{center}
\includegraphics[width=10cm]{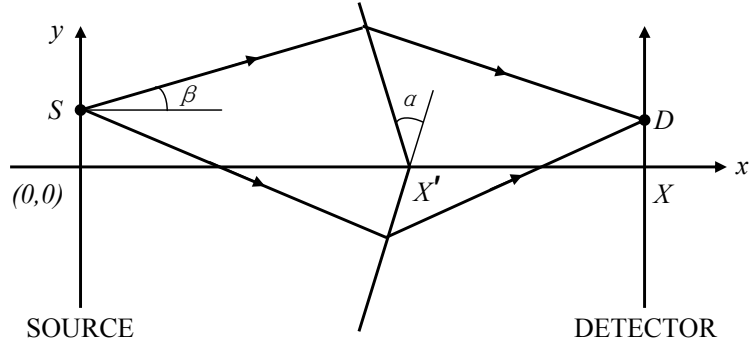}
\caption{Schematic diagram of the simulation setup of a single-photon experiment with a Fresnel biprism.
The apex of the Fresnel biprism with summit angle $\alpha$ is positioned at $(X',0)$.
In the simulation, a line source emits particles from positions
drawn from the current distribution Eq.~(\ref{Jy3}) with random angles $\beta$
chosen uniformly from the interval $[-\alpha/2, \alpha/2]$.
The detectors $D$ positioned at $(X,y)$ count the photons.}
\label{ds_real}
\end{center}
\end{figure}

\begin{figure}[t]
\begin{center}
\includegraphics[width=10cm]{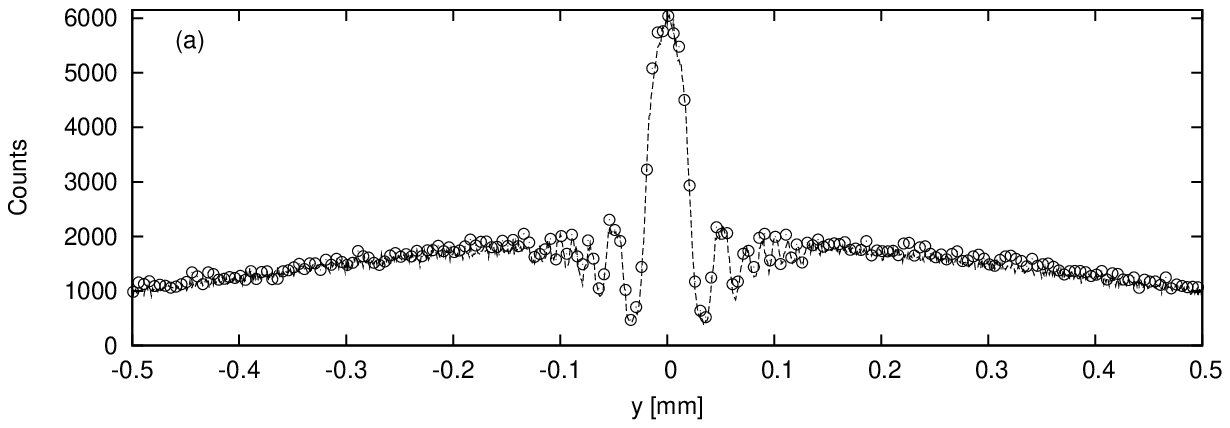}
\includegraphics[width=10cm]{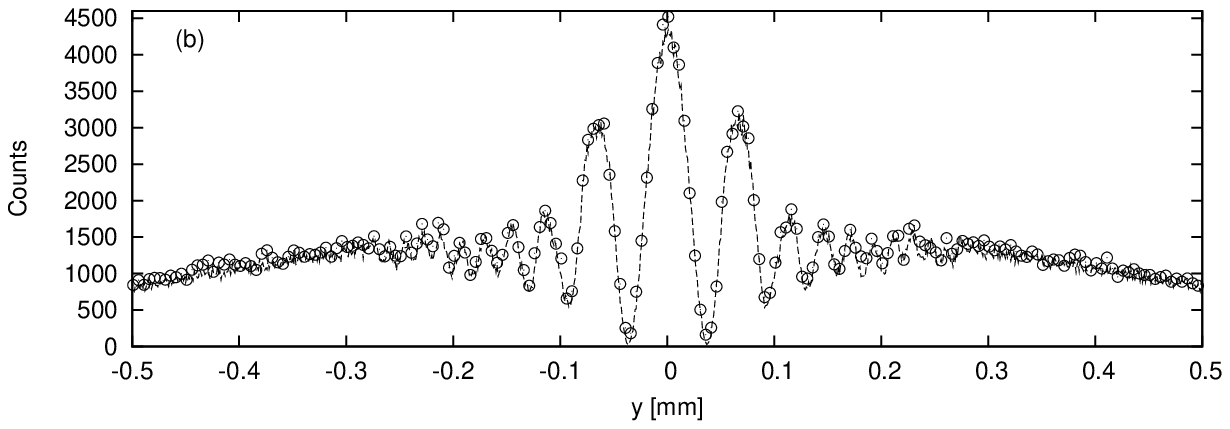}
\includegraphics[width=10cm]{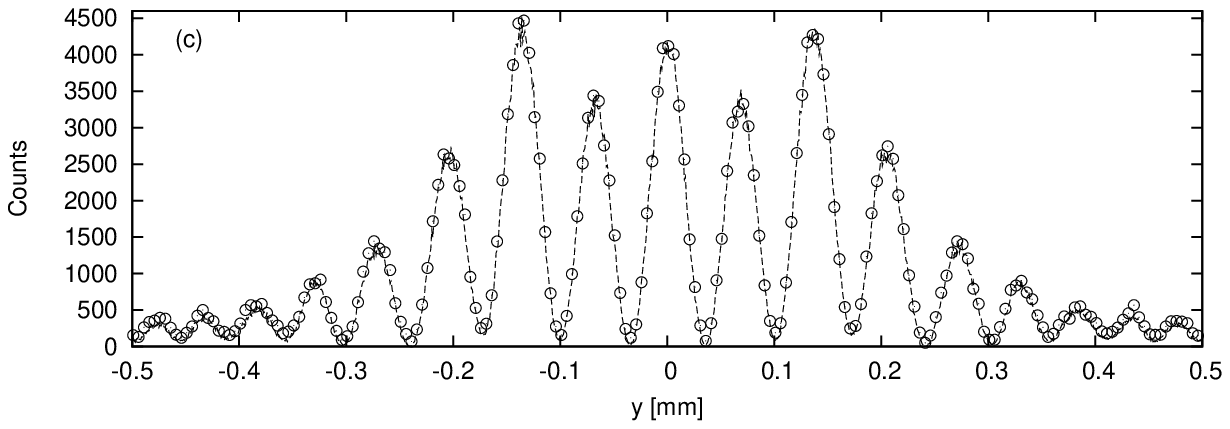}
\caption{Detector counts as a function of the detector position $y$
of the detector array positioned at $X$
for a single photon interference experiment with a Fresnel biprism (see Fig.~\ref{ds_real}).
The Fresnel biprism has an index of refraction $n=1.5631$ and
a summit angle $\alpha=1^{\circ}$. Its apex is positioned at $(X',0)$ with $X'=45$ mm.
The source emits particles according to a Gaussian current distribution with variance $\sigma=0.531$ mm
and wavelength $\lambda=670$ nm~\cite{JACQ05}.
The circles denote the event-based simulation results.
The dashed lines denote the numerical results as obtained from wave theory.
(a) $X-X'=7$ mm; (b) $X-X'=15$ mm; (c) $X-X'=55$ mm. Thousand detectors where used to record the individual events.}
\label{simu22}
\end{center}
\end{figure}

Finally, we consider the single-photon experiment with a Fresnel biprism~\cite{JACQ05}.
Figure~\ref{ds_real} shows the schematic representation of the single-photon interference experiment that we simulate.
For simplicity, we assume that the source $S$ is located in the Fresnel biprism.
Then, the results do not depend on the dimensions of the Fresnel biprism.
Simulations with a Fresnel biprism of finite size yield results that differ
quantitatively only (results not shown).

Messengers are created at positions drawn randomly from the distribution
\begin{equation}
 J(x,y) = \delta(x)e^{-y^2/2\sigma^2}
,
\label{Jy3}
\end{equation}
As in all other cases, the time-of-flight of the messenger
is calculated according to the rules of geometric optics~\cite{BORN64}.
A messenger starting at $(0,y)$ with angle $\beta$
(see Fig.~\ref{ds_real}) leaves the Fresnel biprism at
\begin{eqnarray}
x_\pm&=&\frac{X'\mp y\tan\alpha/2}{1\pm\tan\beta\tan\alpha/2}
,
\nonumber \\
y_\pm&=&\frac{y'+X'\tan\beta}{1\pm\tan\beta\tan\alpha/2}
,
\end{eqnarray}
where the sign has to be chosen such that $x_\pm\le X'$ and $\pm y_\pm\ge0$,
that is such that the path of the messenger crosses the Fresnel biprism boundary.
Using the fact that the tangential component of the velocity
is continuous across the Fresnel biprism boundary~\cite{BORN64}, we have
\begin{eqnarray}
\beta'_\pm=\frac{\pm\alpha}{2}+\arcsin \left[n\sin(\beta\mp\frac{\alpha}{2})\right]
,
\end{eqnarray}
and we find that the messenger hits the screen at $D=(X,(X-x_\pm)\tan\beta'_\pm+y_\pm)$
and that the total time traveled is given by
\begin{eqnarray}
t=n\frac{x_\pm}{c}\sec\beta+\frac{X-x_\pm}{c}\sec\beta'_\pm
.
\end{eqnarray}

In the simulation, the angle of incidence $\beta$ of the photons is selected
randomly from the interval $[-\alpha /2,\alpha /2]$, where $\alpha$ denotes
the summit angle of the Fresnel biprism.
A collection of representative simulation results for $\gamma=0.999$ 
is presented in Fig.~\ref{simu22}.
The dashed lines are the numerical results obtained from wave theory by Monte Carlo sampling.
Again, we find that there is excellent quantitative agreement between
the event-by-event simulation data and wave theory.
Furthermore, the simulation data presented in Fig.~\ref{simu22} is qualitatively very similar to the results
reported in Ref.~\cite{JACQ05} (compare with Fig.~4(d) and Fig.~5(a)(b) of Ref.~\cite{JACQ05}).
Figure 4(c) and (4d) of Ref.~\cite{JACQ05} are made of approximately 20000 photocounts on the CCD camera,
while the number of photodetections on the avalanche photodiodes in absence of the CCD camera would be $40.10^6$ during the exposure
time of 2000 s. Hence the ratio of detected to emitted photons is of the order of 0.0005.
This ratio is much smaller than what we observe in our idealized simulation experiment.
Namely, each of the thousand detectors making up the detection area is hit on average by sixty thousand photons
and the number of clicks generated by the detectors is approximately $16.10^5$.
Hence, the ratio of detected to emitted photons is of the order of 0.026, much larger than the 0.0005 observed
in experiment~\cite{JACQ05}.

\begin{figure}[ht]
\begin{center}
\mbox{
\includegraphics[width=5.5cm]{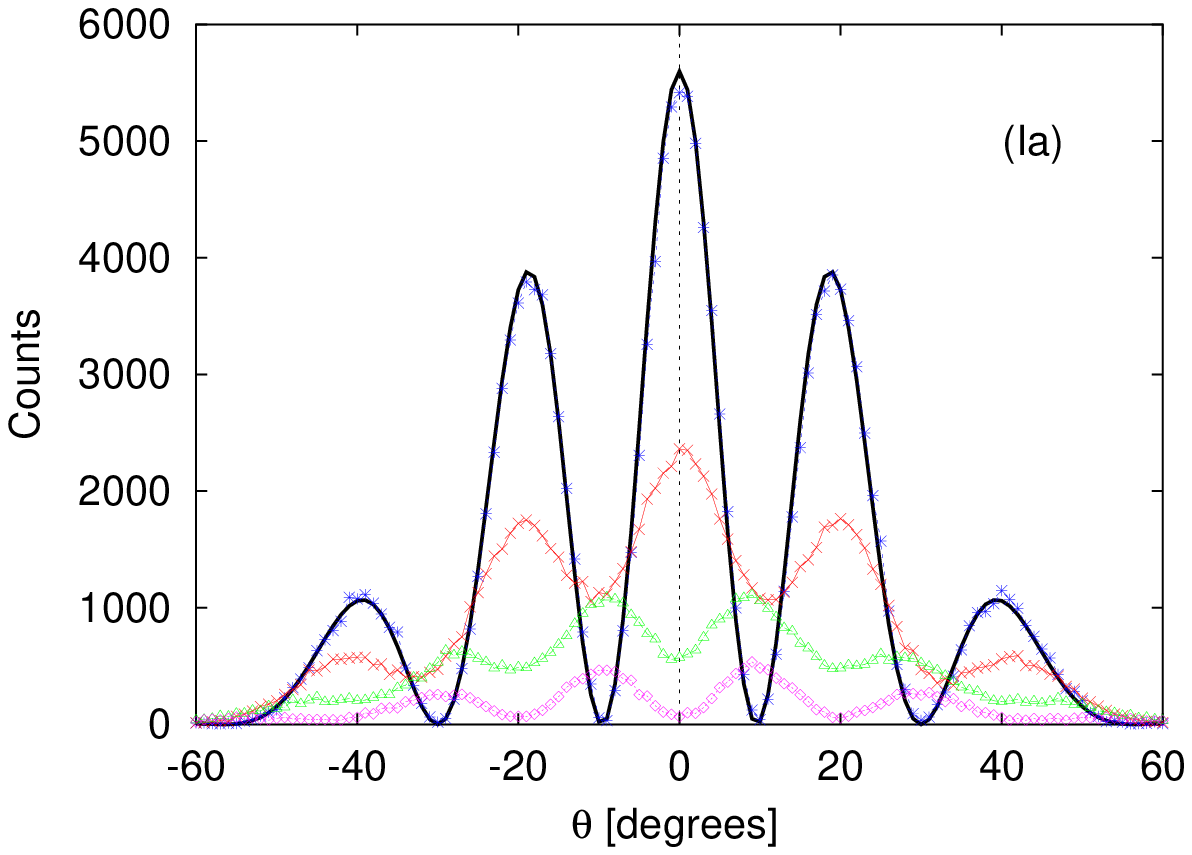}
\includegraphics[width=5.5cm]{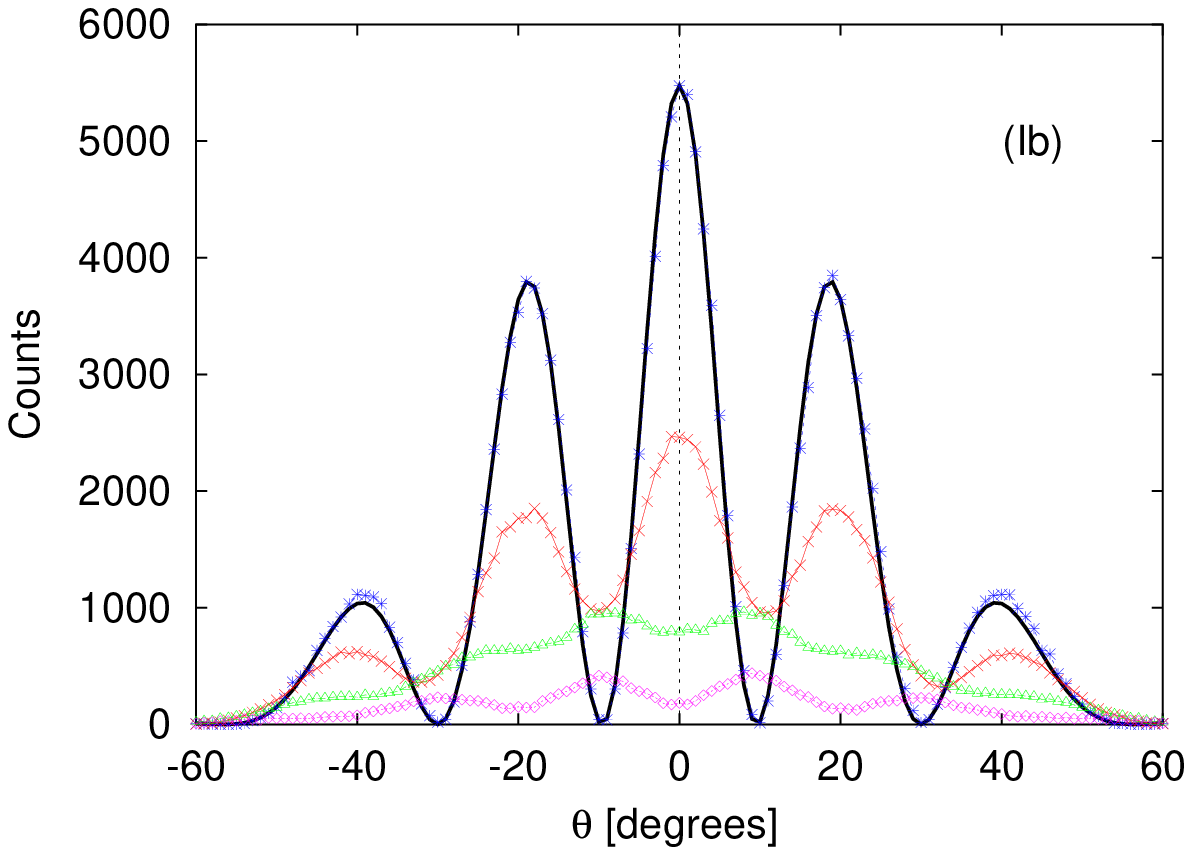}
}
\mbox{
\includegraphics[width=5.5cm]{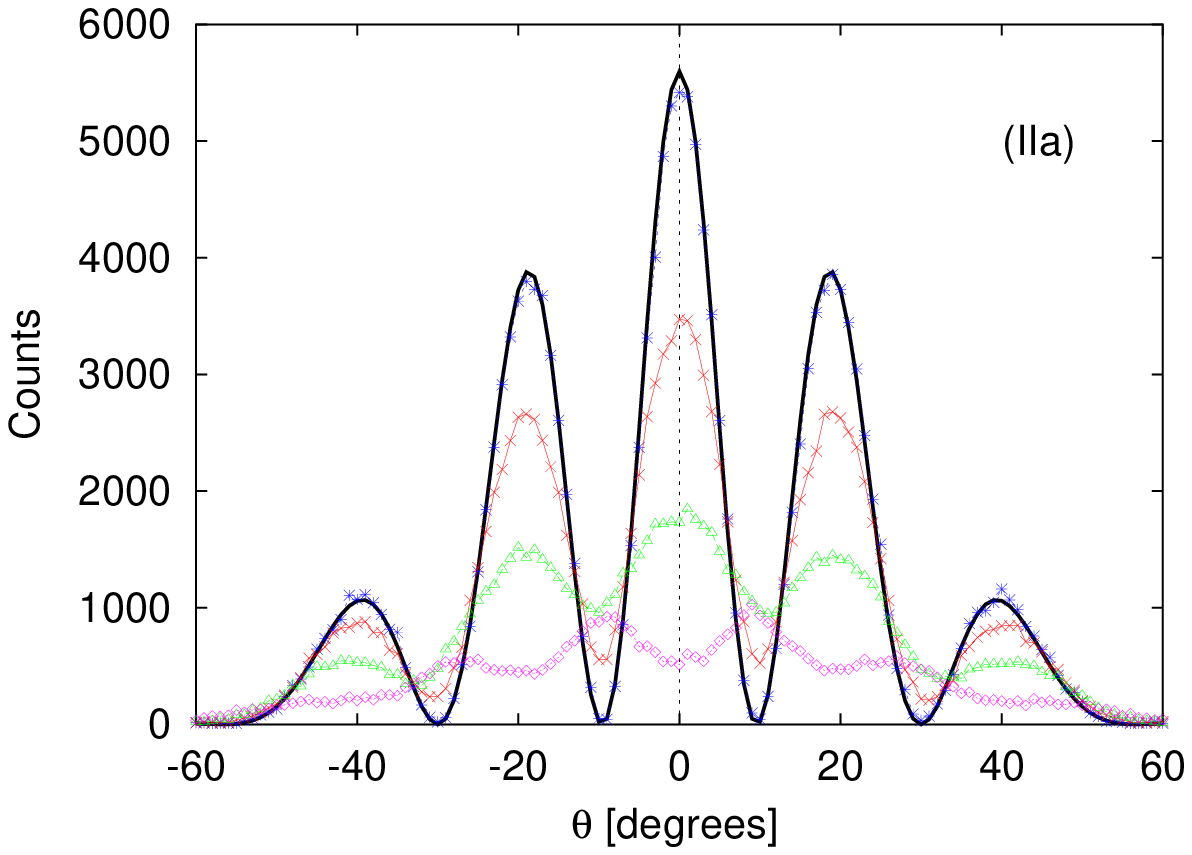}
\includegraphics[width=5.5cm]{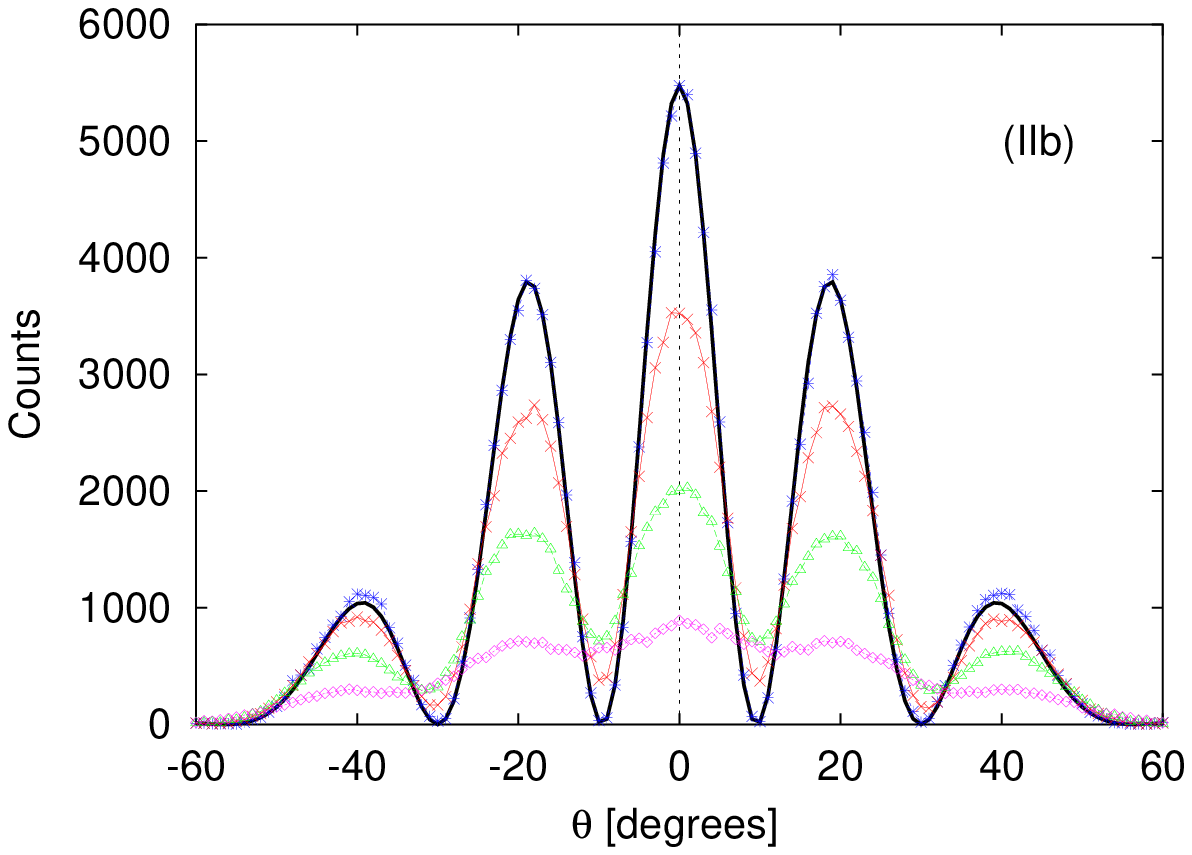}
}
\mbox{
\includegraphics[width=5.5cm]{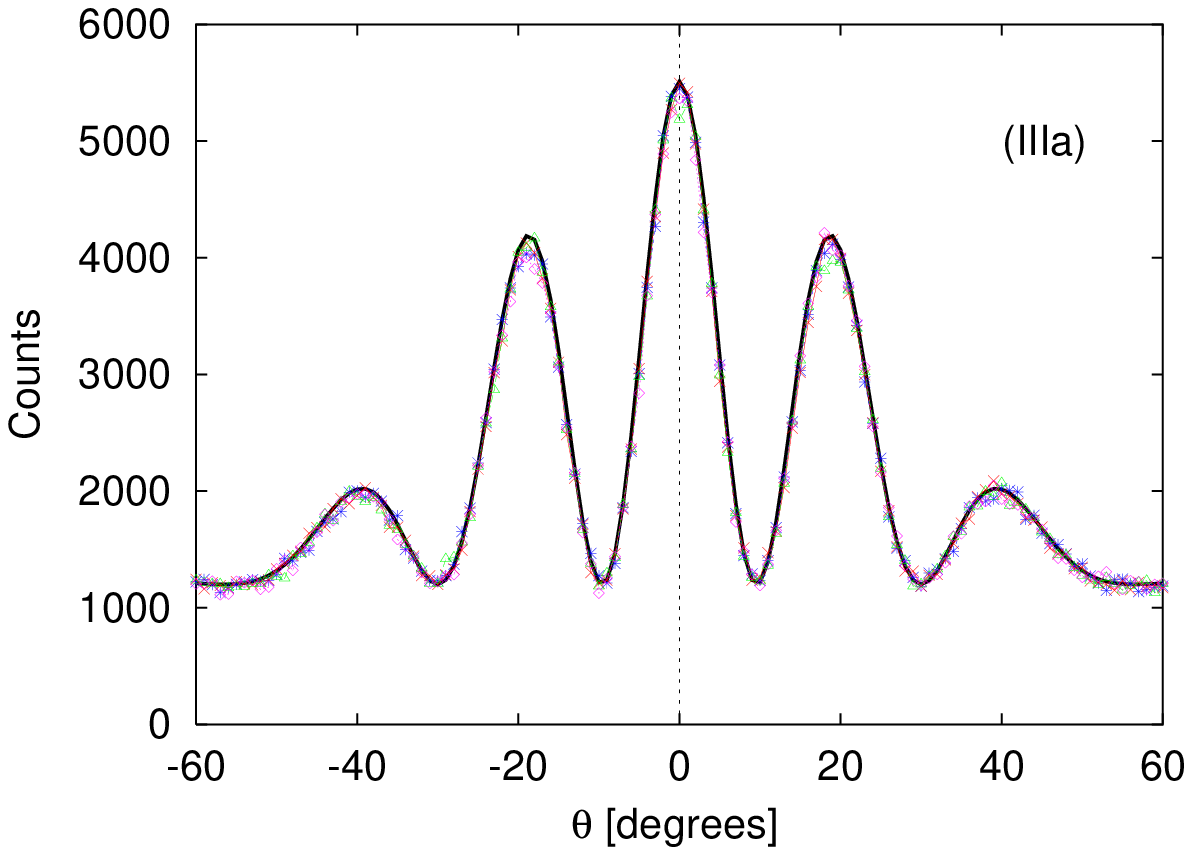}
\includegraphics[width=5.5cm]{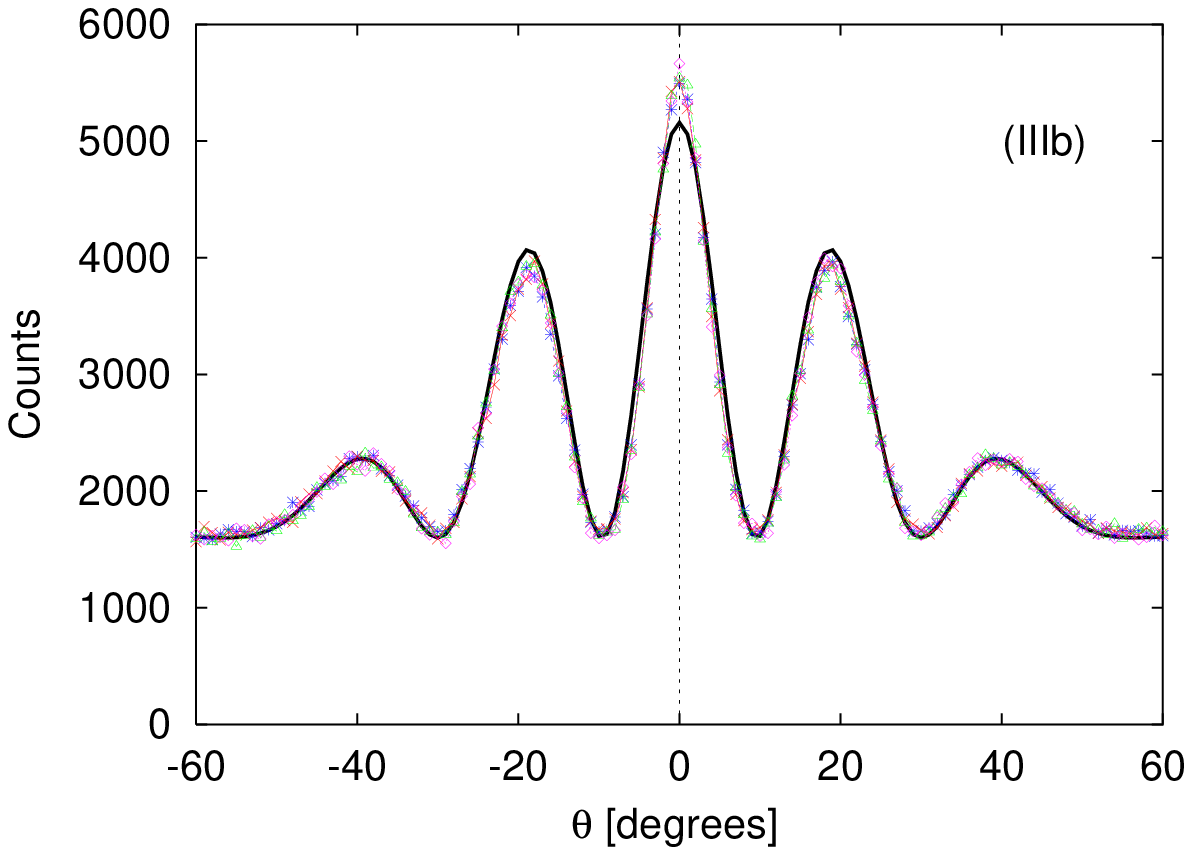}
}
\caption{(color online) Detector counts as a function of the angular detector position $\theta$
for the interference experiment shown in Fig.~\ref{fig2}
which employs only one single-photon detector that is swept over the half circle with a fixed angular velocity.
The results are obtained from event-by-event simulations with six different detector models.
The line sources have a width $a=\lambda$ are separated by a center-to-center
distance $d=3\lambda$, and $X=0.05$ mm (see Fig.~\ref{fig2}).
The labels in the figures indicate the detector model (algorithms) used.
Roman numbers refer to the DLM update rule. The letters a and b refer to the pseudo-random and
deterministic generation of clicks, respectively.
Ia: Eqs.~(\ref{ruleofLM}) and (\ref{thresholdofLM});
Ib: Eqs.~(\ref{ruleofLM}) and (\ref{LM1});
IIa: Eqs.~(\ref{ruleofLM2}) and (\ref{thresholdofLM});
IIb: Eqs.~(\ref{ruleofLM2}) and (\ref{LM1});
IIIa: Eqs.~(\ref{prop2}) and (\ref{thresholdofLM});
IIIb: Eqs.~(\ref{prop2}) and (\ref{LM1}).
Stars: $N_{sweeps}=1$;
Crosses: $N_{sweeps}=25$;
Triangles: $N_{sweeps}=50$;
Diamonds: $N_{sweeps}=100$;
Solid lines: Wave theory, see Eq.~(\ref{eq_slit}).
Other lines are guide to the eye only.
}
\label{prop0}
\end{center}
\end{figure}

\section{Experimental tests: A proposal}\label{proposal}

The simulation models that we propose in this paper make specific
predictions that may be tested by carefully designed,
time-resolved single-photon interference experiments.
However, not all experiments one can think off are as easy to realize.
One of the simplest proposals to test the simulation models would be to consider a
large number ($M$) of identical and independent two-beam (or double-slit) interference
experiments in which only one photon is detected at each of the $M$ detection screens.
According to quantum theory, summing up the single spots of the $M$ detection screens
gives the same interference pattern as if one would conduct one two-beam interference
experiment with $M$ photons being detected on the same detection screen
(all under the assumption that every time a photon is emitted and that all
emitted photons are detected).
For this experiment, the simulation models that we have introduced
do not yield an interference pattern, as is clear from their description.
Thus, at least in principle, this experiment should be able to refute the corpuscular model.
Note that in the absence of any experimental evidence and bearing
in mind that quantum theory has nothing to say about individual events~\cite{HOME97},
it is only a hypothesis that the experiment with finite $M$ will yield results that agree with quantum theory.
Whether this hypothesis is actually true remains to be demonstrated by an experiment.
Unfortunately, in practice, this experiment may be difficult to realize,
the central question being how large $M$ should be before
one observes a pattern that resembles the one predicted by wave theory.
A rough estimate, based on experiments with electrons~\cite{TONO98} suggests that
$M>50000$, a number which makes this proposal very hard to realize in practice.
Therefore, we propose another experiment that may be realizable with
present-day technology.

As explained earlier, if our simulation models operate
in the stationary-state regime, they reproduce the wave theoretical results.
Therefore, to falsify our event-based models
the single-photon experiment should be designed such that
it is sensitive to the transient behavior of the whole setup.
In other words, the experiment should operate on a time scale
that is sufficiently short to prevent the DLM in our detector models
to reach the stationary state.
For a fair comparison between experiment and our simulation models,
it is essential that the experimenter does not discard data
that is recorded during the ``calibration'' or ``warm-up'' stage
because this data may contain valuable information about
the transient behavior of the experimental setup.

In this section, we use our simulation approach
to make predictions of laboratory experiments that may be realizable.
Consider again the double-slit experiment depicted in Fig.~\ref{fig2}
but instead of having many detectors at different angles
$\theta$, we use only one detector placed on a goniometer.
The idea is to keep the total exposure time constant
while the detector is swept
back-and-forth over (part of) the half-circle (see Fig.~\ref{fig2}).
In our simulation models, the recorded interference pattern will then depend on the angular velocity of the detector.
For velocities that are sufficiently small to allow the DLM to reach the stationary state,
the interference pattern obtained agrees with the one predicted by wave theory.
On the other hand, if the detector position changes rapidly,
the DLM may not receive enough events to accurately
reproduce the wave mechanical result.
Therefore, if we keep the total exposure time constant
and perform a set of experiments for several
choices of the sweep velocity,
our simulation models predict that the interference
patterns will change and that these changes reflect
the internal dynamics of the detector model used.

The procedure that we propose is the following.
First, we fix the angle $\delta\theta$ by which the detector position will be moved.
For simplicity, we assume that the aperture of the detector is equal to $\delta\theta$.
Then, we fix the total number of events $N_{total}$
which, on average, will arrive within each arc of angle $\delta\theta$.
Finally, we select the number of times $N_{sweeps}$ that the detector
will be swept back-and-forth over the half circle.

In the simulation, the internal variables of the detector models are initialized once.
The simulation results presented in this section have been obtained
using $\delta\theta=1^\circ$, $N_{total}=10^6$, $N_{sweeps}=1,25,50,100$,
$\gamma=0.999$, and for the modified detector models introduced in Appendix B,
$\kappa=0.9$, $w_0=0.9$ and $\nu=0.99$.
In all figures, the theoretical result Eq.~(\ref{eq_slit}) is rescaled to fit to the maximum of the simulation
data at the smallest sweep velocity and, in the case of IIIa and IIIb, also shifted to account for the non-zero
bias.

As explained in Appendix B, the simple detector model introduced in Section~\ref{model} with
the DLM defined by Eq.~(\ref{ruleofLM}) may require a significant amount (order of thousands)
of input events to reach the stationary state.
Hence, if we move the detector before the DLM reaches its
stationary state, this detector model may not produce results that agree with wave theory.
This expectation is confirmed by the results shown in Figs.~\ref{prop0}(Ia) and (Ib).
If the detector moves slowly ($N_{sweeps}=1$),
the event-based simulation data are in concert with wave theory, as is clear from
the comparison of the 
stars and the 
solid lines in Figs.~\ref{prop0}(Ia) and (Ib).
From Figs.~\ref{prop0}(Ia) and (Ib) it is also clear that
increasing the number of sweeps to $N_{sweep}=25$ (recall that the total amount of events corresponding
to the total exposure time in the experiment is fixed)
leads to a reduction of the visibility of the fringes.
If we increase the number of sweeps to $N_{sweep}=50$, the detector model fails qualitatively.

Thus, an experiment that uses a moving detector might be able to rule out event-based models Ia and Ib
as candidate descriptions of the single-photon interferences.
However, this does not yet imply that our approach as such should be abandoned:
It may be that the detector model is too simple.
Therefore, it is of interest to explore to what extent the results depend on the
particular algorithms used.

It is not difficult to modify the DLM defined by Eq.~(\ref{ruleofLM}) such that
the convergence to the stationary state is much faster or that the response to changes
in the input data is faster.
In Appendix B, we give the details of two of such variants.

DLM II is constructed such that its stationary state behavior is the same as that
of the simple DLM (Eq.~(\ref{ruleofLM})), hence the detector model using this DLM
reproduces the results of wave theory if we employ an array of detectors
or move the single detector very slowly.
From Figs.~\ref{prop0}(IIa) and (IIb),  we may conclude that this model is an improvement over
the simple model in that it still shows interference fringes at a sweeping rate of $N_{sweep}=50$.
For $N_{sweeps}=100$, the detector receives approximately
$N_{total}/(180 N_{sweeps}/\delta\theta)\approx 55$ events
before it moves to the next position.
With this small amount of input events, DLM II
does not reach the stationary state (see also Fig.~\ref{simu1b}).

DLM III is a little different than DLM II: It is sensitive to differences between the internal state
and the input message.
As Figs.~\ref{prop0}(IIIa) and (IIIb) show, these detector models produce output signals that
are insensitive to the speed at which the detector moves but this comes at the price
of a nonzero bias which is, within statistical fluctuations, independent of the detector position
or velocity. Subtracting this bias, all the data fit the theoretical curve very well.

Summarizing: For experiments that use detectors that have fixed positions,
our event-based models for the detector yield results that
cannot be distinguished from those of wave theory.
However, our simulation models for single-photon two-beam interference
show features that may be tested experimentally by measuring the intensity as a
function of the speed of a moving detector.
We have proposed and analyzed a realizable, time-resolved experiment that
directly probes the dynamics of our detector models
and predicted the outcome of such future experiments.

\section{Conclusion}\label{conclusion}

We have demonstrated that it is possible to give a corpuscular description for
single-photon interference experiments with a double-slit, two beams, and with a Fresnel biprism.
Our event-by-event simulation model
\begin{itemize}
\item{does not require any knowledge about the solution of a wave equation,}
\item{reproduces the results from wave theory,}
\item{satisfies Einstein's criterion of local causality,}
\item{provides a simple, logically consistent, particle-based description of interference.}
\end{itemize}

We do not exclude that there are other event-by-event algorithms that
reproduce the interference patterns of wave theory.
For instance, in the case of the single-electron experiment with the biprism~\cite{TONO98},
it may suffice to have an adaptive machine handle the electron-biprism interaction without having
adaptive machines modeling the detectors. We leave this topic for future research.

We hope that our simulation results will stimulate the design of new time-resolved single-photon experiments
to test our corpuscular model for interference.
In Section~\ref{proposal}, we proposed such an experiment and also predicted the outcome
if our simulation model captures the essence of the event-based processes.
Note however that the models we have employed are not unique, as shown explicitly in Section~\ref{model}.
This leaves some freedom to adapt the simulation models to the actual experiments that will be performed.

Finally, it may be of interest to mention that our approach opens a route for
incorporating interference phenomena
into ray-tracing software.
In our simulation method, each messenger simply follows one of the rays through the
medium, updating the message (corresponding to the phase information) as it travels along.
Therefore, for applications where the solution of the Maxwell equations is prohibitive, the combination
of our technique and ray tracing may be a viable alternative.

\section{Acknowledgment}
We would like to thank K. De Raedt, K. Keimpema, S. Zhao, M. Novotny and B. Baten for many helpful comments.
This work is partially supported by NCF, the Netherlands,
by a Grant-in-Aid for Scientific Research on Priority Areas,
and the Next Generation Super Computer Project, Nanoscience Program from MEXT, Japan.

\section*{Appendix A}
We demonstrate that as $\gamma\rightarrow1^-$
the internal vector ${\mathbf p}_{k}$ in Eq.~(\ref{ruleofLM})
converges to the average of the messages ${\mathbf e}_{1},{\mathbf e}_{2},\ldots$.

Let $\Vert{\mathbf x}\Vert$ denote the Euclidean norm of the vector ${\mathbf x}$.
Then, as $0<\gamma<1$, $\Vert{\mathbf e}_k\Vert=1$ for all $k>0$, and $\Vert{\mathbf p}_0\Vert=1$
it follows immediately from Eq.~(\ref{sol0}) that
$\Vert{\mathbf p}_k\Vert\le 1$ for all $k>0$, hence
$\lim_{k\rightarrow\infty} {\mathbf p}_k$ exists.
To determine ${\mathbf p}=\lim_{k\rightarrow\infty} {\mathbf p}_k$, we have to make
assumptions about the properties of the sequence $\{{\mathbf e}_{1},{\mathbf e}_{2},\ldots\}$.
For instance, if the sequence $\{{\mathbf e}_{1},{\mathbf e}_{2},\ldots\}$ is generated by a stochastic process with mean
$\langle{\mathbf e}_{j+1}\rangle={\mathbf e}$ for $j=0,\ldots,k-1$, then it is easy to show that
${\mathbf p}={\mathbf e}$. Thus, in this case, the machine defined by the
rule Eq.~(\ref{ruleofLM}) learns the average ${\mathbf e}$
by updating its internal vector for each message it receives.

In practice, only finite sequences $\{{\mathbf e}_{1},{\mathbf e}_{2},\ldots,{\mathbf e}_{K}\}$
are available. In this case, we can estimate the limiting value by assuming
that the sequence repeats itself, an assumption that is common in Fourier analysis and signal processing
in general~\cite{MANO05}.
From Eq.~(\ref{sol0}), we have
\begin{eqnarray}
{\mathbf p}_{nK}
&=& \gamma^{K} {\mathbf p}_{(n-1)K} + (1-\gamma) \sum_{j=(n-1)K}^{nK-1} \gamma^{nK-j-1}{\mathbf e}_{j+1}
\nonumber \\
&=& \gamma^{K} {\mathbf p}_{(n-1)K} + (1-\gamma) \sum_{j=0}^{K-1} \gamma^{K-j-1}{\mathbf e}_{j+1+(n-1)K}
\nonumber \\
&=& \gamma^{K} {\mathbf p}_{(n-1)K} + (1-\gamma) {\mathbf f}_{K}
,
\label{sol1}
\end{eqnarray}
where
\begin{equation}
{\mathbf f}_{K} = \sum_{j=0}^{K-1} \gamma^{K-j-1}{\mathbf e}_{j+1}
,
\label{sol2}
\end{equation}
and $n>0$. From Eq.~(\ref{sol1}) we find
\begin{equation}
{\mathbf p}_{nK} = \gamma^{nK} {\mathbf p}_{0} + (1-\gamma)\frac{1-\gamma^{nK}}{1- \gamma^K} {\mathbf f}_{K}
,
\label{sol3}
\end{equation}
and hence
\begin{equation}
\lim_{n\rightarrow\infty}{\mathbf p}_{nK} = \frac{1-\gamma}{1- \gamma^K}\sum_{j=0}^{K-1} \gamma^{K-j-1}{\mathbf e}_{j+1}
,
\label{sol4}
\end{equation}
such that
\begin{equation}
\lim_{\gamma\rightarrow1^-}\lim_{n\rightarrow\infty}{\mathbf p}_{nK}
= \frac{1}{K}\sum_{j=0}^{K-1} {\mathbf e}_{j+1}
.
\label{sol5}
\end{equation}
From Eq.~(\ref{sol5}), we conclude that as $\gamma\rightarrow1^-$
the internal vector ${\mathbf p}_{k}$
converges to the average of the messages ${\mathbf e}_{1},\ldots,{\mathbf e}_{K}$.
In general, the parameter $\gamma$ controls the precision
with which the machine defined by Eq.~(\ref{ruleofLM}) learns the average of a sequence of messages
and also controls the pace at which new messages affect the internal state of the learning machine~\cite{RAED05d}.

\section*{Appendix B}
Without performing any simulation, we can already see from Eq.~(\ref{ruleofLM}) that
the simple machine may not perform very well in some cases. Suppose that ${\mathbf p}_{0}=0$ and that
${\mathbf e}_k={\mathbf e}$ for all $k$. Then, from Eq.~(\ref{ruleofLM})
it follows that ${\mathbf p}_{k}=(1-\gamma^k){\mathbf e}$ such that
$\Vert {\mathbf p}_{k}-{\mathbf e}\Vert=\gamma^k$.
Although the latter equation shows that the convergence of ${\mathbf p}_{k}$ to the
input vector ${\mathbf e}$ is exponentially fast, for $\gamma$ very close to
one, in practice, it may take quite a number of events to reach the stationary state.

In this Appendix, we describe two modifications of the algorithm Eq.~(\ref{ruleofLM})
of the first stage (DLM) and one alternative for the algorithm Eq.~(\ref{thresholdofLM}) of the second stage.
The modifications of the first stage reduce the amount of events required for the detector model
to reach the stationary regime.
The alternative for the second stage eliminates the need for a pseudo-random number generator.


It is not difficult to modify the machine such that its asymptotic behavior remains
the same while improving, significantly, the speed with which it learns
from the input ${\mathbf e}_k$.
A simple, but by no means unique, modification is to add one memory element
to store one variable, denoted by $w_{k}$, which keeps track of the differences
between ${\mathbf p}_{k}$ and ${\mathbf p}_{k-1}$.
For $k>0$, these variables are updated according to the rule
\begin{eqnarray}
{\mu}_{k-1} &=& \gamma (1-w_{k-1})
,
\nonumber \\
{\mathbf p}_k &=& \mu_{k-1} {\mathbf p}_{k-1} + (1-\mu_{k-1}) {\mathbf e}_k
,
\nonumber \\
{w}_k &=& \kappa {w}_{k-1} + (1-\kappa)\frac{\Vert{\mathbf p}_{k}-{\mathbf p}_{k-1}\Vert}{2},
\label{ruleofLM2}
\end{eqnarray}
where $0<\kappa<1$ is another control parameter
and $0\le {w}_0\le 1$.
Although the variable ${\mu}_{k}$ is redundant, we wrote
Eq.~(\ref{ruleofLM2}) such that it is obvious that it is an extension
of Eq.~(\ref{ruleofLM}).
In essence, instead of keeping $\gamma$ fixed
in the rule to update ${\mathbf p}_k$ (see Eq.~(\ref{ruleofLM})),
in Eq.~(\ref{ruleofLM2}), the value of $\mu_k$ in the rule to update ${\mathbf p}_k$
is made variable.
This flexibility is then exploited through the first and last rule in Eq.~(\ref{ruleofLM2}).
The last rule defines a machine that learns the distance between
${\mathbf p}_{k}$ and ${\mathbf p}_{k-1}$, the learning speed being
controlled by $\kappa$.
The basic idea is that if this distance is large (say close to but less than $2$),
the last rule will drive $w_k$ to one
such that $\mu_k$ is small and
the change of ${\mathbf p}_{k}$ may be large.
In the opposite situation,
the last rule will force $w_k$ to zero
and ${\mathbf p}_{k}$ will change by small amounts (assuming $\gamma$
is close to but less than one).
As $\mu_k\le\gamma$, the asymptotic behavior of the machine
defined by the rule Eq.~(\ref{ruleofLM2}) is easily shown
to be the same as that of the simple version in which we keep $\mu_k=\gamma$.
Thus, although equations that govern
the dynamics of the machine Eq.~(\ref{ruleofLM2})
are nonlinear (in the ${\mathbf p}$'s), asymptotically the dynamics is
governed by the linear equation Eq.~(\ref{ruleofLM}).

\begin{figure}[t]
\begin{center}
\includegraphics[width=10cm]{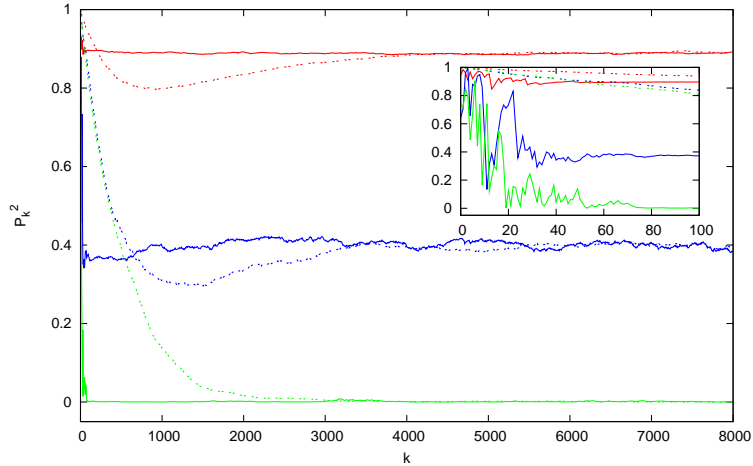}
\caption{(color online) The square of the length of the internal vector ${\mathbf p}^2_{k}$
as a function of the number of received events $k$ for three
different input messages ${\mathbf e}_{k}=(r^{1/2}_k,(1-r_k)^{1/2})$ (top lines),
${\mathbf e}_{k}=(\cos\pi r_k,\sin\pi r_k)$ (middle lines),
and ${\mathbf e}_{k}=(\cos2\pi r_k,\sin2\pi r_k)$ (bottom lines)
where the $0\le r_k<1$ are uniform pseudo-random numbers.
Dashed lines: Model Eq.~(\ref{ruleofLM}).
Solid lines: Model Eq.~(\ref{ruleofLM2}) with $\kappa=0.9$.
The inset shows the short-time response of models Eq.~(\ref{ruleofLM}) and Eq.~(\ref{ruleofLM2}) in more detail.
In all cases ${\mathbf p}_{0}=(1,0)$, $w_0=0.9$ and $\gamma=0.999$.
}
\label{simu1b}
\end{center}
\end{figure}

It is not easy to study the transient behavior of the classical, dynamical systems defined by
Eqs.~(\ref{ruleofLM}) and (\ref{ruleofLM2}) by analytical methods but it is almost trivial
to simulate these models on a computer.
In Fig.~\ref{simu1b}, we show some representative simulation results to illustrate
that the slightly more complicated machine Eq.~(\ref{ruleofLM2})
performs significantly better than the simple machine
Eq.~(\ref{ruleofLM}) with respect to the number of events it takes
for the machine to reach the stationary state.
Roughly speaking, after about 60 events,
machine Eq.~(\ref{ruleofLM2}) has learned enough
to reproduce the correct averages.
As expected on theoretical grounds, both machines
converge to the same stationary state.

A minor modification of algorithm Eq.~(\ref{ruleofLM2}) yields the DLM defined by
\begin{eqnarray}
{\mu}_{k-1} &=& \gamma (1-w_{k-1})
,
\nonumber \\
{\mathbf p}_k &=& \mu_{k-1} {\mathbf p}_{k-1} + (1-\mu_{k-1}) {\mathbf e}_k
,
\nonumber \\
{w}_k &=& \kappa {w}_{k-1} + (1-\kappa)\frac{\Vert{\mathbf p}_{k}-{\mathbf e}_{k}\Vert}{2}
.
\label{prop2}
\end{eqnarray}
Note that the only change is in the third rule where we replaced ${\mathbf p}_{k-1}$
by ${\mathbf e}_{k}$. This replacement causes the machine to respond
very fast to changes in the sequence of input messages $\{{\mathbf e}_{k}\}$
but, at the same time, also leads to a reduction of the average value
of $\mu_{k-1}$ which in turn, will cause the detector model to produce
a nonzero signal, independent of the input messages (see Figs.~\ref{prop0}(IIIa) and (IIIb)).

As an alternative to the pseudo-random ``click generator'' Eq.~(\ref{thresholdofLM}), we may
generate the clicks by means of a very simple DLM~\cite{RAED05b}
containing a single internal variable $0\le z_k\le 1$
that is updated according to
\begin{eqnarray}
       S_k&=&\left\{\begin{array}{ll} 0 & \mbox{if $|{\mathbf p}^2_k-\nu z_{k-1}|< |{\mathbf p}^2_k-\nu z_{k-1}-1+\nu|$}\\
       1 & \mbox{otherwise}\end{array}\right.
       ,
	\nonumber 
       \\
	z_k &=& \nu z_{k-1} + (1-\nu)S_k
       .
	\label{LM1}
\end{eqnarray}
Here, the parameter $0<\nu<1$ plays the same role as $\gamma$ in Eq.~(\ref{ruleofLM}).
The non negative number ${\mathbf p}^2_k$ is the input message for the DLM.
The dynamics of the system defined by Eq.~(\ref{LM1}) is very different
from that of Eq.~(\ref{ruleofLM})~\cite{RAED05b}.
Elsewhere, we have shown that for a fixed input message ${\mathbf p}^2$,
the machine defined by Eq.~(\ref{LM1})
generates a binary sequence (the $S_k$'s) such that in the long run
the ratio of the number of ones relative to the total number of events
is equal to the time average of ${\mathbf p}^2_k$~\cite{RAED05b}.
Thus, the machine defined by Eq.~(\ref{LM1}) produces
clicks with a rate that is determined by ${\mathbf p}^2_k$.

\section*{Appendix C: Relation between simulation model and wave mechanics}\label{relation}

The simulation results presented in Section~\ref{results} demonstrate that the event-based model is
capable of reproducing the results of wave theory without making recourse to the solution of the wave equation
or even a single concept of wave theory.
As there seems to be a general consensus that such models are not supposed to exist,
it is of interest to show that
for the problems that we deal with in this paper,
the event-based model contains the description that derives from Maxwell's equations.

Our demonstration consists of two steps.
First we relate the variables of the event-based model to those of classical electrodynamics.
Second, in analogy with the derivation of the diffusion equation from the discrete random walk model,
we show how our event-based model leads to the Debye model for the interaction
between material and electric field. Other models such as the Drude or Lorentz model can
be derived in a similar manner but to keep the presentation concise,
these derivations are relegated to a future paper.

As is evident from Table~\ref{tab0}, the messenger can be viewed as the event-based equivalent of a classical,
linearly polarized electromagnetic wave with frequency $f$:
The message ${\mathbf e}_k$ corresponds to a plane wave with wave vector ${\mathbf q}$ ($q=2\pi f/c$).
The time-of-flight $t_k$ corresponds to the phase of the electric field.
Adding another clock to the messenger suffices to model
the second electric field component orthogonal to the first one,
and hence the fully polarized plane wave~\cite{ZHAO08b}.
For the systems studied in the present paper including this extra feature,
namely the equivalent of the polarization of the wave, is not necessary and therefore
we confine the discussion to messages that are represented by two-dimensional unit vectors.

\begin{table}[t]
\begin{center}
\caption{%
Correspondence between 
Maxwell's theory and   
the particle-based, event-by-event simulation model.
For simplicity of presentation, we consider
the case of a linearly polarized wave only.
}
\label{tab0}       
\begin{tabular}{@{}lcc@{}}
\hline\hline\noalign{\smallskip}
&Classical electrodynamics & Event-based simulation model
\\ \noalign{\smallskip}\hline\noalign{\smallskip}
Description & wave & particle
\\ \noalign{\smallskip}\hline\noalign{\smallskip}
\multirow{4}{*}{Properties} & oscillator frequency $f$ & oscillator frequency $f$
\\
& direction $\mathbf{q}$ & direction $\mathbf{q}$
\\
& propagation time $t$ & time-of-flight $t_k$
\\
& phase velocity $c$ & velocity $c$
\\ \noalign{\smallskip}\hline\noalign{\smallskip}
Message &
$\mathbf{E}=\mathbf{E_0}\cos(\omega t-\mathbf{q}\cdot\mathbf{r}+\varphi)
$ & $\mathbf{e}_k=(\cos2\pi ft_k, \sin2\pi ft_k)$
\\ \noalign{\smallskip}\hline\noalign{\smallskip}
Material & Polarization $\mathbf{P}(t)$ & Internal vector $\mathbf{p}_k$
\\ \noalign{\smallskip}\hline\noalign{\smallskip}
Interaction& \multirow{2}{*}{
$\mathbf{P}(t)=\int_0^t \chi(u) \mathbf{E}(t-u)du$} &
\multirow{2}{*}{$\mathbf{p}_{k}=\gamma \mathbf{p}_{k-1}+(1-\gamma)\mathbf{e}_{k}$}
\\
with material& &
\\
\noalign{\smallskip}
\noalign{\smallskip}\hline\hline
\end{tabular}
\end{center}
\end{table}

\begin{table}[t]
\begin{center}
\caption{%
Analogy between the derivation of the diffusion equation from the random walk model
and the derivation of one of the constitutive equations in Maxwell's theory
from the discrete model Eq.~(\ref{ruleofLM}) proposed in this paper.
The assumptions that the limiting values
$D=\lim_{\delta\rightarrow0}\lim_{\tau\rightarrow0} \delta^2/2\tau$ and
$\Gamma=\lim_{\gamma\rightarrow1^-}\lim_{\tau\rightarrow0} (1-\gamma)/\tau\gamma$
are nonzero and finite are essential to
obtain a well-defined continuum approximation of the
discrete update rules.
}
\label{tab1}       
\begin{tabular}{@{}l@{}p{10pt}ll}
\hline\hline\noalign{\smallskip}
 &&Random walk&Detector model\\
\noalign{\smallskip}\hline\noalign{\smallskip}
Update rule && $p_{l,k+1}=\frac{1}{2}(p_{l+1,k}+p_{l-1,k})$ & $\mathbf{p}_{k}=\gamma \mathbf{p}_{k-1}+(1-\gamma)\mathbf{e}_{k}$ \\
\noalign{\smallskip}
Length scale: $\delta$ &\multirow{2}{*}{$\Bigg\}$}& \multirow{2}{*}{${p}_{l,k}=p(l\delta,k\tau)={p}(x,t)$} &
$\mathbf{p}_{k}=\mathbf{p}(k\tau)=\mathbf{p}(t)$
\\
Time scale: $\tau$&& &$\mathbf{e}_{k}=\mathbf{e}(k\tau)=\mathbf{e}(t)$
\\
\noalign{\smallskip}
Small $\tau$ &&
${p}_{l,k+1}=p(x,t)+\tau\frac{\partial p(x,t)}{\partial t}+{\cal O}(\tau^2)$
&
$\mathbf{p}_{k-1}=\mathbf{p}(t)-\tau\frac{\partial \mathbf{p}(t)}{\partial t}+{\cal O}(\tau^2)$
\\ \noalign{\smallskip}\noalign{\smallskip}
Small $\delta$
&&
\multicolumn{2}{l}{$p_{l\pm1,k}=p(x,t)\pm\delta\frac{\partial p(x,t)}{\partial x}
+\frac{\delta^2}{2}\frac{\partial^2 p(x,t)}{\partial x^2} +{\cal O}(\delta^3)$
}%
\\ \noalign{\smallskip}\noalign{\smallskip}
Small $\delta$ and $\tau$
&&
$\frac{\partial p(x,t)}{\partial t} \approx \frac{\delta^2}{2\tau}\frac{\partial^2 p(x,t)}{\partial x^2}$
&
$
\frac{\partial \mathbf{p}(t)}{\partial t} \approx -\frac{1-\gamma}{\tau\gamma} \mathbf{p}(t)+\frac{1-\gamma}{\tau\gamma}\mathbf{e}(t)
$
\\ \noalign{\smallskip}
$\lim_{\delta\rightarrow0}\lim_{\tau\rightarrow0}$
&& $\frac{\delta^2}{2\tau} \rightarrow D, \;\; 0< D < \infty$
&
\\ \noalign{\smallskip}
$\lim_{\gamma\rightarrow1^-}\lim_{\tau\rightarrow0}$
& & &
$\frac{1-\gamma}{\tau\gamma}\rightarrow \Gamma,\;\; 0< \Gamma < \infty$
\\ \noalign{\smallskip}
\noalign{\smallskip}
Equation
&& $\frac{\partial p(x,t)}{\partial t}=D\frac{\partial^2 p(x,t)}{\partial x^2}$
&
$\frac{\partial \mathbf{p}(t)}{\partial t} = -\Gamma \mathbf{p}(t)+\Gamma \mathbf{e}(t)$
\\ \noalign{\smallskip}
Fourier space
&
&
$\mathbf{p}(\omega)= \Gamma(i\omega+\Gamma)^{-1} \mathbf{e}(\omega)$
\\ \noalign{\smallskip}
&&
&
$\mathbf{P}(\omega)= \chi(\omega) \mathbf{E}(\omega)$
\\ \noalign{\smallskip}
&&
~~~~~~~~~~~~~$\Downarrow$
&
~~~~~~~~~~~~~$\Downarrow$
\\
&&Diffusion equation
&Constitutive equation
\\
\noalign{\smallskip}
\noalign{\smallskip}\hline\hline
\end{tabular}
\end{center}
\end{table}

The internal vector ${\mathbf p}_k$ plays the role of the polarization vector ${\mathbf P}(t)$
of the detector material.
Indeed, comparing the formal solution of Eq.~(\ref{ruleofLM})
\begin{equation}
{\mathbf p}_{k} = \gamma^k {\mathbf p}_{0} + (1-\gamma) \sum_{j=0}^{k-1} \gamma^{j}{\mathbf e}_{k-j}
,
\label{sol0}
\end{equation}
with the constitutive equation
\begin{equation}
{\mathbf P}(t)=\int_0^t \chi(u) {\mathbf E}(t-u) du
,
\label{sol0a}
\end{equation}
in Maxwell's theory~\cite{BORN64}, it is clear that both equations have the same mathematical structure:
The left hand sides are convolutions of the incoming (applied) message (field)
with memory kernel $\gamma^j$ ($\chi(u)$) (in applications, we may assume that the initial value ${\mathbf p}_{0}=0$).
Thus, the DLM is a simple model for the interaction of the individual photons with
the material of the detector. The time-of-flight, corresponding to the phase
of the electric field, is used to update the internal vector which corresponds
to the polarization vector of the material.

Next, we show that this analogy can be carried much further
by mimicking the derivation that relates the discrete
random walk on a line to the one-dimensional diffusion equation~\cite{GRIM95}.
The essential steps for both the random walk and our event-based
detector model are summarized in Table~\ref{tab1}.
Both models describe a process that proceeds in discrete time steps $\tau$.
The random walk model is formulated on a lattice with mesh size $\delta$.
In the case of the random walk, we let the time step $\tau$ and mesh size $\delta$ go to zero.
In the event-based model we let the time step $\tau$, that is the time between the arrival
of successive messages, approach zero and let $\gamma$ approach one.
For both models, we demand that the resulting continuum equations make sense.
This enforces relations between $\tau$ and $\delta^2$ and between $\tau$ and $\gamma$,
as shown in Table~\ref{tab1}.
Then, the former relation yields an explicit expression of the diffusion coefficient $D=\delta^2/2\tau$
in terms of the length and time scale of the discrete random walk model.
Likewise, the latter leads to the Debye model for a dispersive medium~\cite{TAFL05} and gives
an explicit expression for the relaxation time $1/\Gamma=\tau\gamma/(1-\gamma)$ in terms of the
parameters of the event-based model.

As Table~\ref{tab1} shows, under certain conditions, the discrete models can be approximated by continuum equations
that describe the coarse-grained (in space-time for the random walkers and in time for the event-based model)
behavior but the discrete models provide a description with details that can never be extracted from
the corresponding continuum equations.
Of course, the ultimate justification of the event-based model is that,
as shown in Section~\ref{results}, it can reproduce the results of wave theory.
Appendix D gives a further justification of our approach from a computational point of view.

\section*{Appendix D: Computational point of view}\label{problem}
There is a general consensus that unless we first solve the wave equation
and use this solution as the probability distribution for generating events,
there are very fundamental, apparently unsurmountable, problems
to derive from a wave mechanical description a process that produces the events that are observed
in experiment~\cite{HOME97}.
The arguments used are rather abstract and general~\cite{HOME97} and
to understand the subtilities that are involved it may help
to address this issue from a computational point of view.

For phenomena that cannot (yet) be described by a deductive theory, it is common practice to use probabilistic models.
Although Kolmogorov's probability theory provides a rigorous framework to formulate such models,
there are ample examples that illustrate how easy it is to make plausible assumptions that create all kinds of paradoxes,
also for every-day problems~\cite{GRIM95,TRIB69,JAYN03,BALL03}.
Subtle mistakes such as dropping (some of the essential) conditions,
like in the discussion of the double-slit experiment~\cite{BALL86,BALL01},
mixing up the meaning of physical and statistical independence or
changing one probability space for another during the course of an argument, can give rise to
all kinds of paradoxes~\cite{JAYN89,HESS01,BALL86,KHRE01,BALL03,HESS06}.
For instance,
Feynman used the double-slit experiment as an example to argue that ``far more fundamental was the discovery that in nature
the laws of combining probabilities were not those of the classical probability theory of Laplace''~\cite{FEYN65b},
but this statement has been shown to result from an erroneous application of probability theory~\cite{BALL86,BALL01,BALL03}.

By construction, if we use a digital computer to produce numbers
as we do in this paper, we stay in the domain of elementary arithmetic and we do not have to worry about
the subtleties of Kolmogorov's probability theory.

Instead of discussing the apparently unsurmountable problem in its full generality, which we could,
it is more instructive to examine in detail the simple, concrete example of
the double-slit model depicted in Fig.~\ref{fig2}.
According to Maxwell's theory, in the Fraunhofer regime
the light intensity at the detector on a circular screen is given by~\cite{BORN64}
\begin{eqnarray}
\frac{I(\theta)}{I(0)}&=&
\left\vert
\int_{-\infty}^{+\infty}
e^{iqy'\sin\theta}\rho(y') dy'
\right\vert^2
,
\label{fun0}
\\
&=&\left(\frac{\sin\frac{qa\sin\theta}{2}}{\frac{qa\sin\theta}{2}}\right)^2 \cos^2\frac{qd\sin\theta}{2}
,
\label{fun1}
\end{eqnarray}
where $\rho(y')=[\Theta(a-|y'-d/2|)+\Theta(a-|y'+d/2|)]/2a$
is the normalized density distribution for the coordinate $y'$.

First, starting from the explicit expression Eq.~(\ref{fun1})
for the density $I(\theta)/I(0)$,
it is trivial to construct an algorithm that generates events
according to this density.
Indeed, let us define
\begin{equation}
S_j(\theta)=\Theta(I(\theta)-r_j I(0))
,
\label{fun2}
\end{equation}
where $0\leq r_j <1$ denotes a uniform pseudo-random number.
Then, the number of clicks of the detector at angular position $\theta$ is
given by
\begin{equation}
N_k(\theta)=\frac{1}{k}\sum_{j=1}^k S_j(\theta)
,
\label{fun3}
\end{equation}
and for sufficiently large $k$, we have $N_k(\theta)\rightarrow I(\theta)/I(0)$ with probability one.
This completes the construction of the event-based algorithm based on the knowledge of $I(\theta)/I(0)$.
Obviously, this algorithm is built on the knowledge of the explicit solution $I(\theta)$ of the wave problem.
The events generated by this algorithm build up the interference pattern one-by-one
and can be identified with the clicks of the detectors.
This is as far as the quantum theoretical description goes in making contact to the experimental observations:
It provides a prescription to calculate the probability density to observe a click on a detector.
It is quite common to postulate that there does not exist a description
that goes beyond the specification of the probability, excluding
that no further advance in a deeper understanding of the process that produces the events can be made.

Disregarding this postulate, we may wonder what happens if we take one step back
and assume that we only know about expression Eq.~(\ref{fun0})
in terms of the wave amplitudes $\exp(iqy'\sin\theta)$ and density $\rho(y')$.
Then, the obvious thing to do is to compute the integral in Eq.~(\ref{fun0}) numerically.
Without loss of generality, we may write
\begin{eqnarray}
A(\theta)&=&
\frac{1}{N({\cal S})}\sum_{y'\in {\cal S}}e^{iqy'\sin\theta}
,
\label{fun4}
\end{eqnarray}
where the summation is over all $y'$ of the set ${\cal S}$
accounting for the density $\rho(y')$ and $N({\cal S})$ is the normalization factor.
By definition of the integral, if the number of elements of the set ${\cal S}$ goes to infinity,
we have $\left\vert A(\theta)\right\vert^2\rightarrow I(\theta)/I(0)$.

Although the numerical calculation of the amplitude $A(\theta)$
is straightforward, there obviously is no relation
between the points $y'$ of the set ${\cal S}$ and the number of clicks of the
detector at $\theta$.
In fact, the essence of quantum theory is that there is only a relation between
$\left\vert A(\theta)\right\vert^2$ and the number of clicks
but to know $A(\theta)$ requires that we first generate (a lot of) pseudo-events $y'$.
Obviously, these pseudo-events $y'$ cannot have an interpretation in terms of
observed clicks.

The conclusion therefore is that the description in terms of individual waves (Eq.~(\ref{fun0}))
does not contain the ingredients, not even conceptual,
to define a process that generates the clicks of the detectors that we observe.
Therefore, from a computational perspective, it is futile to try inventing an event-based, particle-like process
based on the wave mechanical expression for the intensity in terms of sums over amplitudes.

One may take the position that it is fundamentally impossible to
go beyond an event-level description based on the knowledge
of $I(\theta)/I(0)$ but by postulating this to be true, one simply postulates that
it is impossible to make any advance in a deeper understanding
of event-based phenomena. As we have shown by this and many earlier papers,
there is no rational argument that supports this postulate other than that
it is what we have been taught in physics courses.

Having shown that our event-by-event simulation model
reproduces the results of wave theory without resorting to
a description in terms of waves, we now explain why,
from a computational point of view, we consider this
to be an accomplishment and why our approach works.

The crux of our approach is that we do not start
from expression Eq.~(\ref{fun0}) but construct
a discrete event process that converges to Eq.~(\ref{fun0})
while generating events that directly correspond to the observed events.
During the initial phase, this process may generate events that
are accidental but once the process has reached
its stationary state, the events appear with
frequencies that corresponds to those predicted by wave theory.

To understand the idea behind our approach,
it may be helpful to draw an analogy with
the well-known Metropolis Monte Carlo (MMC) method
for solving statistical mechanical problems~\cite{HAMM64,LAND00}.
The MMC method generates states $S$, events in our terminology,
with a probability density~\cite{HAMM64,LAND00}
\begin{equation}
p(S)=\frac{e^{- E(S)/k_BT}}{\sum_S e^{-E(S)/k_BT}}
,
\label{sub2}
\end{equation}
where $E(S)$ denotes the energy of the state $S$,
$k_B$ is Boltzmann's constant and $T$ is the temperature.
At first sight, sampling from Eq.~(\ref{sub2}) is impossible
because in all but a few nontrivial cases for which the
partition function $\sum_S e^{-E(S)/k_BT}$ is known,
we do not know the denominator.
The MMC method solves this problem by constructing a Markov chain that generates
a sequence of events $S$ such that asymptotically these events are distributed according
to the (unknown) probability density Eq.~(\ref{sub2})~\cite{HAMM64,LAND00}.

The analogy with our approach is the following.
Although very different in all details, our event-based method uses a deterministic process
(implemented as a DLM, see Eq.~(\ref{ruleofLM})) of which the sampling distribution
converges to the unknown probability distribution $I(\theta)/I(0)$.
The one-to-one correspondence between the objects in the corpuscular, event-based description
and those in Maxwell's theory (see Section~\ref{relation})
ensures that in the long run, the event-based detector model generates clicks
with frequencies that correspond to those of the unknown probability distribution $I(\theta)/I(0)$.

\bibliographystyle{jpsj}
\bibliography{../../../epr}   

\end{document}